\documentclass{emulateapj}
\usepackage{amsmath}
\begin{document}

\title{Type-Ia SUPERNOVA REMNANT SHELL AT $z=3.5$ SEEN IN THE THREE SIGHTLINES TOWARD THE GRAVITATIONALLY LENSED QSO B1422+231 \footnote{Based on data collected at Subaru Telescope, which is operated by the National Astronomical Observatory of Japan.}}
\author{Satoshi Hamano, Naoto Kobayashi}
\affil{Institute of Astronomy, University of Tokyo, 2-21-1 Osawa, Mitaka, Tokyo 181-0015, Japan; hamano@ioa.s.u-tokyo.ac.jp}
\author{Sohei Kondo}
\affil{Koyama Astronomical Observatory, Kyoto-Sangyo University, Motoyama, Kamigamo, Kita-Ku, Kyoto 603-8555, Japan}
\author{Takuji Tsujimoto}
\affil{National Astronomical Observatory of Japan}
\affil{Department of Astronomical Science, The Graduate University for Advanced Studies, 2-21-1 Osawa, Mitaka, Tokyo 181-0015, Japan}
\author{Katsuya Okoshi}
\affil{Faculty of Industrial Science and Technology, Tokyo University of Science, 102-1 Tomino, Oshamanbe, Hokkaido, 049-3514, Japan}
\and
\author{Toshikazu Shigeyama}
\affil{Research Center for the Early Universe, University of Tokyo,  7-3-1 Hongo, Bunkyo, Tokyo 113-0033, Japan}

\begin{abstract}

Using the Subaru 8.2m Telescope with an IRCS
Echelle spectrograph, we obtained high-resolution ($R=$10,000) near-infrared (1.01-1.38$\mu$m) spectra of images A and B of the gravitationally lensed QSO
B1422+231 ($z=3.628$) consisting of four known lensed images. We detected \ion{Mg}{2} absorption lines at
$z=3.54$, which show a large variance of column densities ($\sim$0.3 dex)
and velocities ($\sim$10 km s$^{-1}$) between the sightlines A and B
with a projected separation of only $8.4h_{70}^{-1}$ pc at the
redshift. This is the smallest spatial structure of the high-$z$ gas
clouds ever detected after Rauch et al. found a 20-pc scale
structure for the same $z=3.54$ absorption system using optical
spectra of images A and C. The observed systematic variances imply
that the system is an expanding shell as originally
suggested by Rauch et al. By combining the data for
three sightlines, we managed to constrain the radius
and expansion velocity of the shell ($\sim$50-100 pc, 130 km s$^{-1}$),
concluding that the shell is truly a supernova remnant
(SNR) rather than other types of shell objects, such as a giant \ion{H}{2}
region.  We also detected strong \ion{Fe}{2} absorption lines for
this system, but with much broader Doppler width than that of $\alpha$-element lines. We
suggest that this \ion{Fe}{2} absorption line originates in a
localized \ion{Fe}{2}-rich gas cloud that is not completely mixed with plowed ambient interstellar gas
clouds showing other $\alpha$-element low-ion absorption lines.
Along with the Fe richness, we conclude that the SNR is produced by an SNIa explosion.

\end{abstract}


\section{Introduction}

Gravitationally lensed QSOs provide us precious opportunities to study spatial structures of high-$z$ gas clouds which intersect the multiple sightlines toward the QSO \citep[e.g.,][]{wey83,fol84,sme95,rau99,rau01a,rau01b,rau02,kob02,chu03a,ell04,lop05,mon09}.
By comparing the profiles of absorption lines between multiple sightlines, we can study the density and velocity gradients of intersecting gas clouds directly even at high redshift. 
Such spatial properties of absorbing gas at high redshift may provide a key to understand the galaxy formation processes.

The statistical studies of absorption line systems with the gravitationally lensed QSOs have revealed that the \ion{Mg}{2} systems, which trace low-ionization systems are relatively small ($<$ a few hundred pc) and have clumpy spatial structures \citep{rau02,ell04}, in contrast to the \ion{C}{4} systems which are typically a few kpc in size and have fewer structures \citep{rau01a,ell04}. Studies of single line-of-sight observations and CLOUDY photoionization modeling \citep{fer98} also suggest that the Fe-rich low-ionization systems should be as small as 10 pc \citep{rig02,nar08}. Thus, low-ionization systems appear to have small clumpy spatial structures that directly relate to star forming activities, such as giant molecular clouds. 
It is important to investigate their spatial properties with multiple sightlines of gravitationally lensed QSOs, since only the gravitational lens can directly reveal fundamental parameters such as the size and kinematics of gas clouds.

The quadruple gravitationally lensed QSO B1422+231 \citep{pat92} is one of the best objects for such a study because of the relatively small separations among multiple sightlines due to the large distance between the lense galaxy \citep[$z=0.339$; ][]{kun97,ton98} and the QSO ($z=3.628$). This QSO has been observed frequently as one of the most luminous gravitationally lensed high-$z$ QSOs \citep{jil95,son96,pet98,rau99,rau01a,rau01b}.
\citet[][, hereafter RSB99]{rau99} obtained high-resolution optical spectra of images A and C of B1422+231 using Keck HIRES \citep{vog94} and detected \ion{C}{4}, \ion{Si}{4}, \ion{C}{2}, \ion{Si}{2}, and \ion{O}{1} absorption lines from the $z=3.54$ absorption system originally identified by \citet{son96} with strong Ly$\alpha$ and Ly$\beta$ absorption lines. The projected separation between A and C sightlines is only 22.2$h^{-1}_{70}$pc at the redshift. RSB99 suggested that the absorption system is an expanding shell of mass ejection or supernova remnants (SNRs) by analyzing the differences of the absorption profiles of lower ionization species between the images A and C. 

\begin{figure}[t]
 \centering
 \includegraphics[width=7cm,clip]{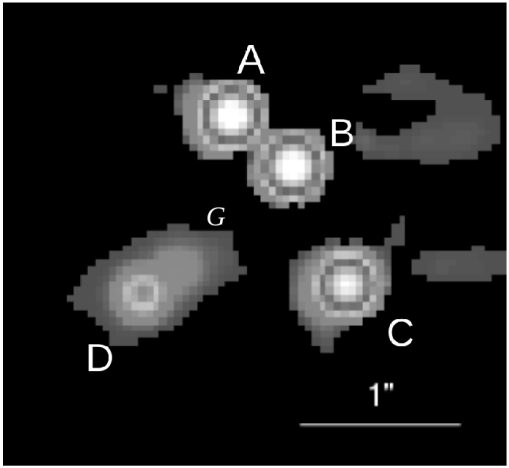}
 \caption{\textit{HST} F555W image of B1422+231 from CASTLES homepage (http://www.cfa.harvard.edu/castles). North is up, east is to the left. Four white point sources show gravitationally lensed images A, B, C and D. The extended object (G) to the northwest of image D shows the lensing galaxy of this system. We obtained spectra of the two brightest and closest components, A and B.}
 \label{img}
\end{figure}

In this paper, we report the results of the Subaru near-infrared spectroscopic observations of images A and B of B1422+231.
Previous studies of low ionization gas with \ion{Mg}{2} QSO absorption lines were limited to the optical wavelength range and thus to redshifts $<2.5$, beyond which this transition moves into the infrared waveband.
Thanks to the advent of highly sensitive near-infrared high-resolution spectroscopy with 8-10 m class telescope, it is possible now to study this unexplored redshift range.
We are conducting a systematic high-resolution spectroscopic survey of high-$z$ \ion{Mg}{2} systems with the Subaru IRCS Echelle spectrograph.
We observed B1422+231 as one of the initial targets, and detected
\ion{Mg}{2} doublet and \ion{Fe}{2} absorption lines for the $z=3.54$ system. Moreover, we succeeded in spatially resolving spectra of images A and B owing to the high spatial resolution in the infrared and the very good seeing of the Subaru Telescope site. This paper is composed as following. The details of our observations are summarized in \textsection 2. The data reduction and calibration of spectra are described in \textsection 3. In \textsection 4, we show the properties of detected \ion{Mg}{2} and \ion{Fe}{2} absorption lines. We interpret the properties as signatures of a Type Ia supernova (SN Ia) remnant, which is discussed in \textsection 5 in detail. \textsection 6 is the summary of this paper.
We adopt a standard cosmology with $\Omega _\Lambda = 0.7$, $\Omega _m = 0.3$, $\Omega _k = 0$ and $H_0 = 70 h_{70} = 70 $ km s$^{-1}$ Mpc$^{-1}$ throughout this paper.

\section{Observation}
\label{Observation}

\begin{figure*}
 \centering
 \includegraphics[width=12cm,clip]{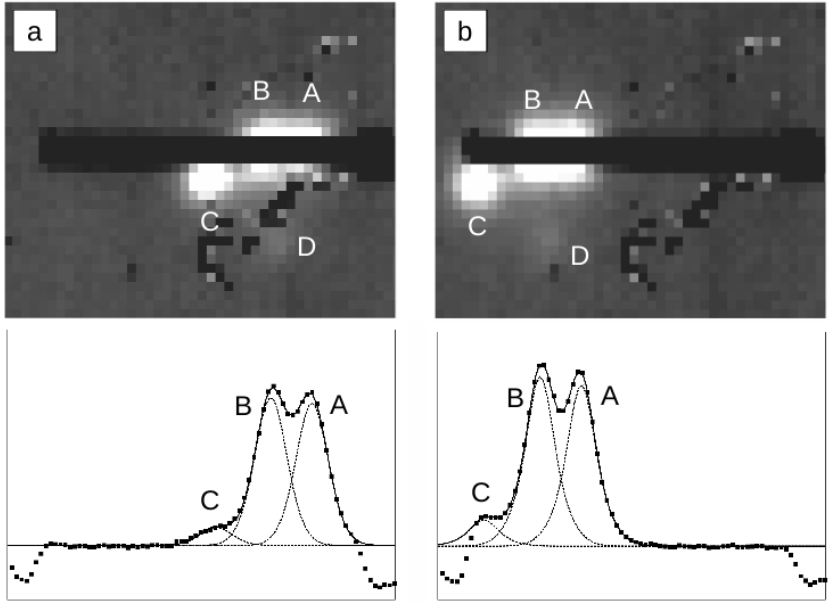}
 \caption{Upper panels: the slitviewer $K$-band images obtained during the echelle observation. These are raw images with an integration time of 120 s without corrections for hot/bad pixels. The pixel scale is $0^{''}.075$ pixel$^{-1}$. The horizontal black thick bar near the center of the images is the shadow of the slit, and four point sources which are seen in white are the images A, B, C and D of B1422+231. The left and right panels correspond to the dither patterns ``a'' and ``b'' (see the text for detail). The contamination of image C in the slit is slightly seen in this image. Lower panels: the spatial profiles of the spectra along the slit. Points show the observed data. Dotted lines show fitted profiles for each image A, B, and C, while thick lines show the combined profile of the three images. To show the spatial profile clearly, the data are averaged in the dispersion axis over one cross-disperser order (order 44 of $J$ band). The hollows at both edges show the areas beyond the end of the slit, which are not used for the profile fitting.}
 \label{slit}
\end{figure*}

We observed images A and B of B1422+231 (Figure \ref{img}) using Subaru 8.2m Telescope \citep{iye04} with IRCS\footnote{IRCS was mounted at the Cassegrain focus at that time. Now it is located at the Nasmyth focus.} Echelle spectrograph \citep{tok98,kob00} in $Y$ (1.01$-$1.19$\mu$m) and $J$ (1.16$-$1.38$\mu$m) bands. The data were obtained on 2003 February 13 and 2002 April 28 for $Y$ and $J$ bands, respectively. The weather condition for both observing runs  was photometric and the seeing was good ($\sim 0^{''}.5$) during observations for both bands. The integration time per frame was 600 sec, and the total integration time was 9000 s and 9600 s for $Y$ and $J$ bands, respectively. The slit widths were  $0^{''}.60$ and $0^{''}.30$, and corresponding spectral resolutions ($R=\lambda / \Delta \lambda$) were $5,000$ and $10,000$ for $Y$ and $J$ bands, respectively, while the slit length was $3^{''}.47$ for both observations. Although we used the Subaru Adaptive Optics 36-elements (AO 36) system\footnote{The AO36 system is now upgraded to AO188 system with much improved image correction capability.} \citep{tak04} for the $Y$-band observation, the improvement of image quality was not good enough that we used the wider slit.
The slit position angle (P.A.) was set at P.A. = $39^\circ_.2$ to put both A and B images on the slit simultaneously. The telescope was nodded by $1^{''}.5$ arcsec along the A$-$B direction as shown in Figure \ref{slit} between alternating frames to offset the sky emission in the subsequent data reduction. We call the frames for positions ``a'' and ``b'' as shown in Figure \ref{slit}. We also observed a standard star, 10 Boo, for flux calibration and removal of telluric absorption lines for both bands. The airmass of the target was distributed in a wide range from 1.0 to 1.5 for both bands, while that of the standard star was about 1.0 and 1.5 for $Y$ and $J$ bands, respectively. 

\section{Data Analysis}

\subsection{Reduction}
\label{Reduction}

We used IRAF\footnote{IRAF is distributed by the National Optical Astronomy Observatory, which is operated by the Association of Universities for Research in Astronomy, Inc., under cooperative agreement with the National Science Foundation.} for data reduction following standard procedures. First, we subtracted ``b'' frames from corresponding ``a'' frames to cancel out the dark counts and the sky OH emission lines. All the subtracted frames were combined after the flat-fielding correction and hot/bad pixel correction. Then, we extracted two-dimensional (2D) spectra with the spatial axis along the slit and the dispersion axis perpendicular to the slit for each cross-disperser order using IRAF task ``apall'' in the ``echelle'' package. The 2D spectra of each order were combined after applying the wavelength calibration with Argon lamp spectra, which were obtained after the observation.

\subsection{Deconvolution of A and B Spectra}

The lower panel of Figure \ref{slit} shows the spatial profile of the 2D spectra of B1422+231 along the slit. The images A and B are almost resolved, but the overlap is not negligible. Image C, which is not a target of this study, is contaminated slightly in the spatial profiles. To obtain precise one-dimensional (1D) spectra of images A and B from the 2D spectra, we fitted the observed spatial profile with the point spread functions (PSFs) of the three images pixel by pixel in the dispersion axis. This method is similar to that of \citet{lop05}, who also observed the multiple images of a gravitational-lensed QSO in a slit simultaneously.

Because the tail of PSF was found to decline slower than that of a Gaussian function, we assumed the following double Gaussians as the form of a PSF:
\begin{equation}
 f_i(x,\lambda) = \frac{a_i(\lambda)}{2} \left(\exp \Big [-\frac{(x-b_i)^2}{\sigma _s^2} \Big] + \exp \Big [-\frac{(x-b_i)^2}{\sigma _d^2} \Big]\right), \label{PSF}
\end{equation}
where $i=$A, B, C are the indices for the lensed images, $x$ is the coordinate of pixels in the spatial axis, $a_i(\lambda)$ is the peak value of each image at wavelength $\lambda$, $b_i$ is the center position of each image, and $\sigma_s$ and $\sigma_d$ are the width of each Gaussian function ($\sigma_s < \sigma_d$ ; the subscripts ``s'' and ``d'' mean sharp and diffuse, respectively). This function was found to fit the observed profile precisely as shown in the lower panel of Figure \ref{slit}. 

The parameters $b_i$, $\sigma_s$, and $\sigma_d$ have to be fixed for linear fitting, which is necessary to converge the fit for observed profiles with low signal to noise ratio (S/N). These parameters are successfully determined from the fit of the high S/N spatial profile made by summing the spatial profiles for 50 pixels along the dispersion axis. The ``fit'' command of gnuplot was used for the fitting procedure. Though this program is not designed for numerical fitting, it is good enough to conduct the linear fitting.
The flux density count at wavelength $\lambda$ for image $i$ was determined by the integral of the PSF $f_i(x,\lambda)$ along the spatial axis, which can be analytically calculated as $\sqrt{\pi} a_i(\lambda) ( \sigma _s + \sigma_d)/2$ for double Gaussians. When all the fitting parameters are determined for the entire wavelength range, the final 1D spectrum of each image is constructed by plotting the flux density count along the dispersion axis. 

\begin{figure*}
	\centering
	\includegraphics[width=14cm,clip]{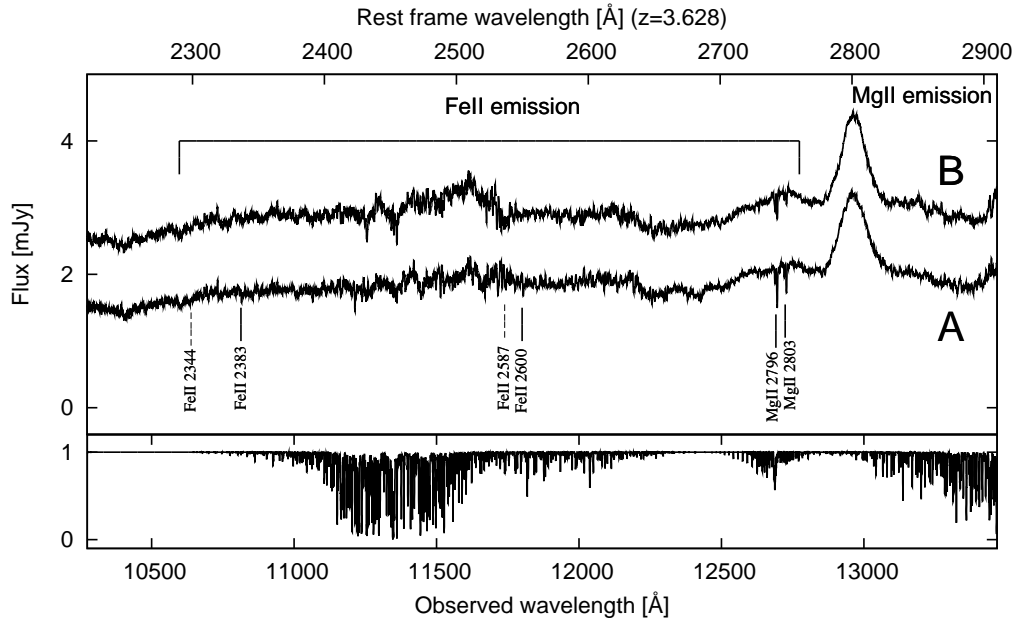}
	 \caption{Upper panel: the whole $Y$- and $J$-band spectra of images A (lower spectra) and B (upper spectra) of B1422+231. The observed spectra were telluric-corrected and smoothed with an 11 pixel ($\sim 3$\r{A}) box car to clearly show the spectra with an effective spectral resolution of R$\sim$4,000. The bumps on the continuum are broad \ion{Mg}{2} and \ion{Fe}{2} emission lines from the QSO itself at $z=$3.628. The vertical bars show the expected positions of absorption lines associated with the $z=3.54$ system: solid and dashed lines show the positions of the detected and non-detected lines, respectively. Lower panel: the telluric absorption spectrum in the same wavelength range. This is made by normalizing the spectrum of the standard star, 10 Boo, after removing two hydrogen absorption lines (Pa$\beta$ and Pa$\gamma$) with Voigt profile fitting.}
 \label{zentai}
\end{figure*}

\begin{figure*}
 \centering
 \includegraphics[width=13cm,clip]{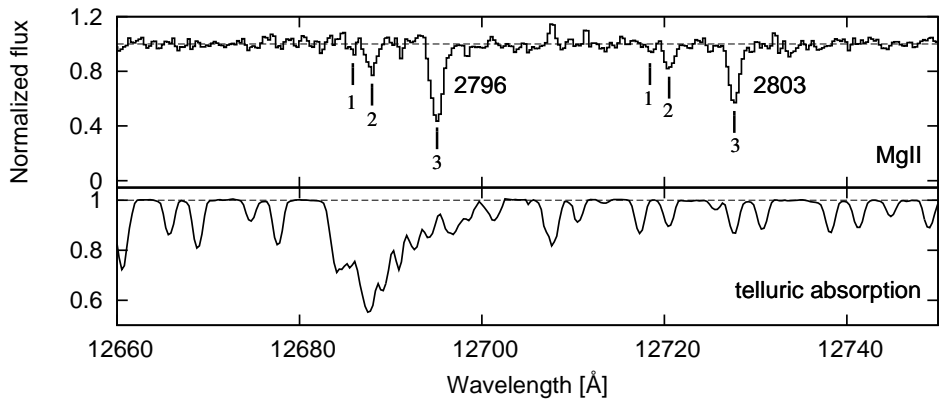}
 \caption{Upper panel: the normalized spectrum of image A around \ion{Mg}{2} doublet lines of the $z=3.54$ system after the telluric corrections. Lower panel: the normalized spectrum of the standard star 10 Boo, showing the telluric absorption lines. The strong O$_2$ absorption band at around 12690 \r{A} apparently affects the detection and profiles of components 1 and 2 of \ion{Mg}{2} $\lambda$2796.}
 \label{telluric}
\end{figure*}

\subsection{1D Spectra}
\label{1Dspectra}

Many telluric absorptions of e.g., water (H$_2$O) and oxygen (O$_2$) plus emission lines of OH appear on the near-infrared spectra for ground-based observations. Although the emission lines are removed by the substraction of images of alternative pointing (\textsection \ref{Reduction}), the absorption lines are still superposed on the extracted spectra of the object. For both bands, we removed the telluric absorption lines by dividing the object spectra by that of a A0 standard star, 10 Boo, which has few intrinsic absorption lines except for two strong hydrogen absorption lines : Pa$\gamma$ ($\sim$ 10935 \r{A}) in $Y$ band (order 51 and 52) and Pa$\beta$ ($\sim$ 12815\r{A})  in $J$ band (order 44). Before the division, we removed these hydrogen lines by fitting with Voigt profile, and scaling the telluric absorption lines to correct the airmass difference (\textsection \ref{Observation}). Figure \ref{zentai} shows the flux calibrated spectra of images A and B. We used the continuum of the telluric  standard star 10 Boo (5.67 mag in $J$-band from SIMBAD\footnote{}) for flux calibration, assuming the effective temperature of 9480 K. 

To avoid the influence of many bumps due to \ion{Fe}{2} and \ion{Mg}{2} \textit{emission} lines from the QSO itself on the continuum fitting, the normalized absorption spectra were made by fitting the continuum in a velocity range of $\pm 500$ km s$^{-1}$, which is narrower than the velocity widths of the emission lines ($\sim 1000$ km s$^{-1}$), around the detected absorption lines using spline3 function. 

Heliocentric corrections of $-$21.89 km s$^{-1}$ and $-$5.18 km s$^{-1}$ were applied to $Y$- and $J$-band spectra, respectively. By comparing the detected \ion{Mg}{2} absorption lines with \ion{C}{2} and \ion{Si}{2} absorption lines for image A (RSB99), a residual offset of $-$3.4 km s$^{-1}$ (about a half pixel) was found in our $J$-band spectrum. We shifted the absorption line spectrum by +3.4 km s$^{-1}$ to match the wavelength of the \ion{C}{2}/\ion{Si}{2} absorption lines with those derived by RSB99 because their spectral resolution ($R=$70,000, $\Delta v = 4.4$ km s$^{-1}$) is much better than ours ($R=$10,000, $\Delta v = 30$ km s$^{-1}$). This offset is likely to come from the slight change of the instrument setting that occurred between the observation of targets and the acquisition of the Argon lamp spectrum. For $Y$-band data, no absorption lines that have obvious peaks as \ion{Mg}{2} absorption lines of the $z=3.54$ system were detected. Therefore, we did not shift the spectra after the heliocentric correction.

\section{$z=3.54$ System}

\begin{deluxetable}{ccccc}
\tabletypesize{\scriptsize}
\tablecaption{Detected \ion{Mg}{2} absorption lines. }
\tablewidth{0pt}
\tablehead{
\colhead{Image} & \colhead{Comp} & \colhead{$W_r$($\lambda$2796)(\r{A})} & \colhead{$W_r$($\lambda$2803)(\r{A})} & \colhead{Doublet ratio}
}
\startdata
 A & 1&  (0.020)&       0.018   & (1.1) \\
 &  2&	(0.074)&	0.052	& (1.4)  \\
 & 3&	0.191	&	0.124	& 1.5 \\
B & 2&	(0.080)&	0.040	& (1.6) \\
 &  3&	0.111	&	0.056	& 2.0 
\enddata
\tablecomments{$W_r$ of the components that are significantly affected by telluric absorption lines are shown with parentheses.}
\label{ew}
\end{deluxetable}

\subsection{Velocity Components}

In the final $J$-band spectrum, strong \ion{Mg}{2} $\lambda \lambda$2796, 2803
absorption lines with two velocity components at $z=3.54$ are clearly
detected for both images A and B (Figure \ref{zentai}). The
corresponding \ion{Fe}{2} $\lambda$2383 ($f=0.320$) and $\lambda$2600 ($f=0.2394$) absorption lines of this system
are also detected in image A (Figure \ref{zentai}), while other \ion{Fe}{2} absorption lines, such as $\lambda$2344 ($f=0.1142$) and $\lambda$2587 ($f=0.0691$), are not detected probably because of their small oscillator strengths. Hereafter, we
will focus only on this $z=3.54$ \ion{Mg}{2} system: the search for other systems and their
results will be discussed separately in S. Kondo et
al. (in preparation). Unfortunately, the \ion{Mg}{2} $\lambda$2796 line is
overlapping with the telluric O$_2$ band as shown in Figure \ref{telluric}
and the systematic uncertainty may be left in the profile of
\ion{Mg}{2} $\lambda$2796 even after the removal of the telluric
absorption lines (\textsection 3.3). The impact of
the telluric absorption lines will be discussed in the following
subsection (\textsection 4.2) for each velocity component. 

\begin{figure}
 \centering
 \includegraphics[width=8cm,clip]{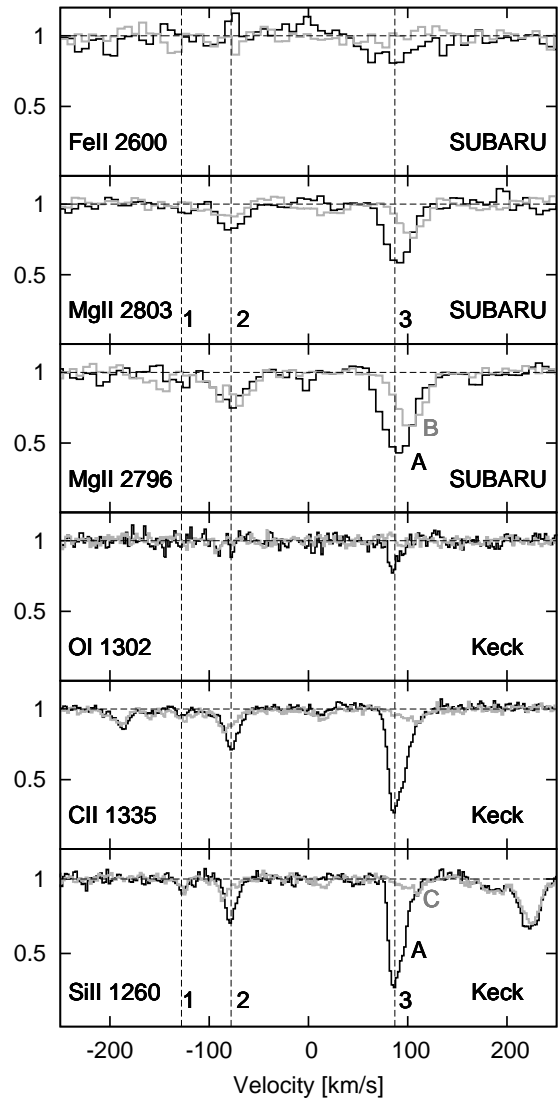}
 \caption{Spectra of the $z=3.54$ absorption system of images A (black lines), B (gray lines for \ion{Mg}{2} and \ion{Fe}{2}), and C (gray lines for \ion{Si}{2}, \ion{C}{2} and \ion{O}{1}) of B1422+231. The spectral resolutions are about 10,000 and 70,000 for our SUBARU data (top three panels) and RSB99's Keck data (bottom three panels), respectively. The vertical dashed lines show the peak position of the three velocity components of image A. The velocity is relative to $z=3.53850$. The atomic species and their rest-frame wavelengths are shown at the left bottom corner of each plot.} 
 \label{spectra}
\end{figure}

This $z=3.54$ system was first detected by \citet{son96} as an
\ion{H}{1} system with $N$(\ion{H}{1}) = $2.3 \times 10^{16} $cm$^{-2}$ in their high-resolution
optical spectrum with Keck HIRES. Later RSB99 detected various metal
absorption lines also with Keck HIRES in spatially-separated spectra
of images A and C. Figure \ref{spectra} shows the \ion{Mg}{2} absorption
lines of the $z=3.54$ system for both images A and B from our $J$-band
data, and \ion{C}{2}, \ion{Si}{2}, and \ion{O}{1} absorption lines of
images A and C from RSB99's optical data. 

\begin{figure*}
 \centering
 \includegraphics[width=7cm,clip]{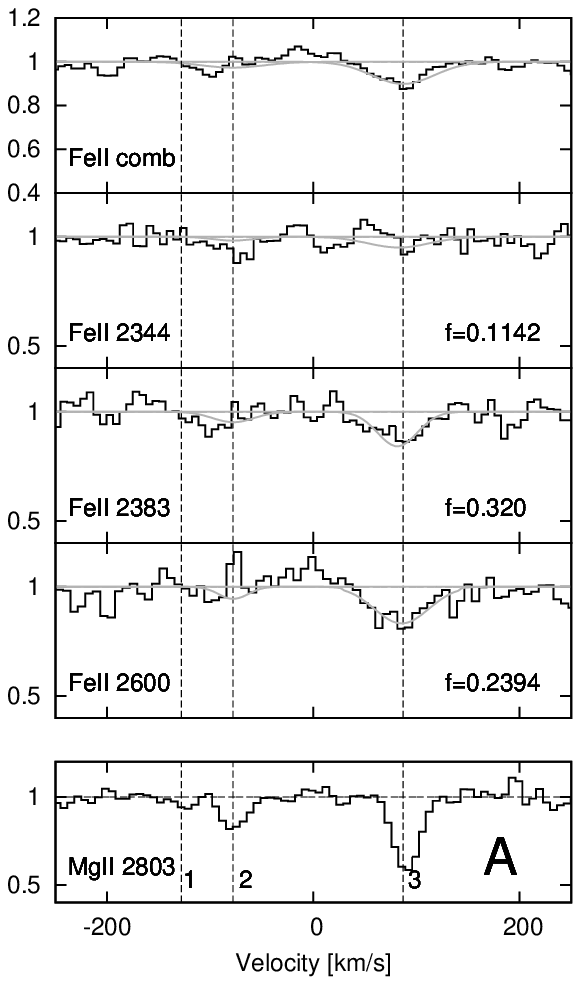}
 \includegraphics[width=7cm,clip]{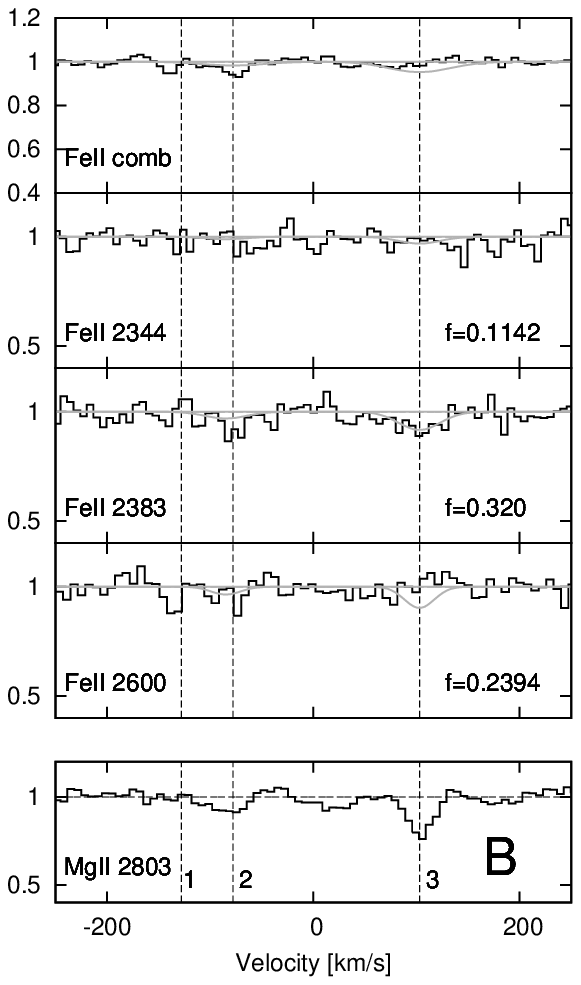}
 \caption{Spectra of \ion{Fe}{2} lines of the $z=3.54$ system for images A and B. The spectrum of  \ion{Mg}{2} $\lambda$2803 line is also shown at the bottom as a reference. The top spectrum ``\ion{Fe}{2} comb'' shows the combined spectra of \ion{Fe}{2} $\lambda$2344, $\lambda$2383 and $\lambda$2600. The species and the rest-frame wavelength are shown at the bottom left of each spectrum. The oscillator strength of each line from \citet{mor91} is shown at the bottom right of each plot. The spectral resolution of \ion{Fe}{2} $\lambda$ 2600 and \ion{Mg}{2} $\lambda$2803 spectra is $R\sim 10,000$, while that of \ion{Fe}{2} $\lambda$2383, 2344 and ``\ion{Fe}{2} comb'' is $R\sim 5000$. The vertical dashed lines show the center positions of three velocity components of each images. The gray lines show the \ion{Fe}{2} absorption profiles predicted from the \ion{Mg}{2} profiles (see the detail in Section 4.3).}
 \label{Fecomb}
\end{figure*}

For the \ion{Mg}{2}
absorption lines, we clearly detected two velocity components for both
images A and B at $\sim -80$ km s$^{-1}$ and $\sim +90$ km s$^{-1}$,
which are identified as ``component 2'' and ``component 3'' in
RSB99. The \ion{Mg}{2} absorption line that corresponds to ``component
1'' is also detected at $\sim -130$ km s$^{-1}$, but only for image A
(Figure \ref{spectra}).  The rest-frame equivalent widths ($W_r$) of detected
\ion{Mg}{2} $\lambda \lambda$2796, 2803 absorption lines are
summarized in Table \ref{ew}. Note that the equivalent widths of
lines that are affected by the strong telluric absorption lines are
shown with parentheses. The total rest-frame equivalent width of $\lambda2796$ for this system is calculated as 0.28\AA. Therefore, this system is classified as a ``weak \ion{Mg}{2} system'' which is defined as a system with $W_r (2796) < 0.3$ \AA \ \citep{chu99}. The corresponding \ion{Fe}{2} $\lambda2383, \lambda2600$
absorption lines are also detected for component 3 but only for image A as
shown in Figure \ref{Fecomb} (left panel). 
The \ion{Mg}{1} absorption line is not
detected for any component for both images within the
uncertainty. 

We fit a Voigt profile to the \ion{Mg}{2} and \ion{Fe}{2} absorption
lines and measured the column density, the relative velocity, and the
Doppler width using VPFIT\footnote{VPGUESS is a graphical interface to
  VPFIT written by Jochen Liske; see
  http://www.eso.org/~jliske/vpguess.}\citep{car87} and
VPGUESS\footnote{VPFIT is a Voigt profile fitting package provided by
  Robert F. Carswell; see http://www.ast.cam.ac.uk/~rfc/vpfit.html.}
software assuming each velocity component is composed of just one velocity sub-component, which is a good approximation for estimating column densities from our data with the velocity resolution of $\Delta v = 30$km s$^{-1}$ at most. The results are summarized in Table \ref{vpfit}. The shown
uncertainty is only that for the VPFIT fitting. For non-detected
absorption lines, we calculated the $3\sigma$ upper limit of the
column density from the S/N of the continuum around the
absorption lines. We discuss the characteristics of the detected
\ion{Mg}{2} and \ion{Fe}{2} absorption lines in the following.

\subsection{\ion{Mg}{2} Absorption Lines}

Table \ref{ew} summarizes the \ion{Mg}{2} doublet ratio,
which is defined as $W_{r}(2796)$/$W_{r}(2803)$, for each component
of each image. Where both $\lambda \lambda$2796, 2803
absorption lines are not saturated, the doublet ratio should be equal
to 2 by simply reflecting the oscillator strengths. For the strongest component (component 3), the doublet ratio is found
to be 1.5 for image A, while equal to 2 for image B. Therefore, the absorption line of image A may be slightly saturated while that of image B is not. Although the
other weaker components are also unlikely to be saturated in view of the
smaller $W_{r}(2803)$ than that for the component 3 of image B, the
doublet ratios for those components show values less than 2. This is
probably because $W_{r}(2796)$ of components 1 and 2 is
underestimated due to the incomplete removal of the heavy telluric
absorption lines that are overlapped on those components (Figure
\ref{telluric}).  Here, we discuss details of the \ion{Mg}{2}
absorption lines for each component 1, 2, and 3 .

\noindent
\textit{Component 1.} For this component, RSB99 shows that there is no
difference between the absorption lines of images A and C, thus the same
absorption profile was expected for image B, which is located in
between images A and C on the sky. However, we could not detect component 1
in image B despite its existence in the spectrum of image A (Figure
\ref{spectra}). In particular, we had expected to detect the \ion{Mg}{2} $\lambda 2803$ absorption line because it is not affected by the strong O$_2$ telluric absorption lines unlike the \ion{Mg}{2} $\lambda 2796$ absorption line. The \ion{Mg}{2} $\lambda$2803
absorption line for component 1 is located right in between two weak
telluric absorptoin lines (Figure \ref{telluric}) and should not be affected by
the removal of the telluric absorption lines.  Therefore, we conclude
that the gas cloud of component 1 covers only images A and C on the
sky, suggesting a small-scale sub-structure on this high-redshift
cloud.

\begin{deluxetable*}{cccccccc}
\tabletypesize{\scriptsize}
\tablecaption{The Results of Voigt Profile Fitting with VPFIT}
\tablewidth{0pt}
\tablehead{
\colhead{Species} & \colhead{Component} & \colhead{$\log N_A\text{(cm}^{-2}\text{)}$} & \colhead{$\log N_B\text{(cm}^{-2}\text{)}$}& \colhead{$v_A$(km s$^{-1}$)} & \colhead{$v_B$(km s$^{-1}$)} &\colhead{$b_A$(km s$^{-1}$)} &\colhead{$b_B$(km s$^{-1}$)}
}
\startdata
  \ion{Mg}{2} & 1 & 11.9$\pm$1.9 & $\cdots$ & -127$\pm$16  & $\cdots$ & 2$\pm$55  & $\cdots$ \\
  & 2 & 12.4$\pm$0.1 & 12.20$\pm$0.07 & -78$\pm$5 & -85$\pm$3 & 7.2$\pm$1.3\footnote{Estimated from the Doppler width of \ion{C}{2} and \ion{Si}{2}. See the detail in Section 4.1}  & 10 $\pm$ 9 \\
  & & (12.5$\pm$ 0.4) & & (-77$\pm$6) & & (5$\pm$29) & \\
   & 3 & 13.03$\pm$0.07 & 12.66$\pm$0.04& 90.6$\pm$0.6& 102.9$\pm$0.6 & 9$\pm$1 & 7$\pm$1\\
  \ion{Fe}{2}  & 2 & $<12.1$ & $<12.1$ & $\cdots$ &$\cdots$ & $\cdots$ &$\cdots$\\ 
   & 3 & 12.72$\pm$0.05 & $<12.1$&83$\pm$3  & $\cdots$ & 23 $\pm$6  & $\cdots$ \\
   \ion{Mg}{1} & 2 & $<11.1$ & $<11.1$&$\cdots$ &$\cdots$ &$\cdots$ &$\cdots$
\enddata
\tablecomments{The column densities, relative velocities from $z=$3.53850, and Doppler widths measured with VPFIT are shown for both line of sight A and B.}
\label{vpfit}
\end{deluxetable*}

\noindent
\textit{Component 2.} Although we tried to fit both \ion{Mg}{2}
$\lambda \lambda$2796, 2803 lines of component 2 simultaneously, the
uncertainties of the three fitting parameters (column density, redshift and Doppler width) were found to be very
large and we could not get any reliable estimate. This is most likely
due to the strong telluric O$_2$ band on the \ion{Mg}{2} $\lambda$2796
line (Figure \ref{telluric}): systematic uncertainties may be left in the
profile of \ion{Mg}{2} $\lambda$2796 even after the removal of
telluric absorption lines (\textsection 3-3), which is consistent with
the strange doublet ratios for component 2 described above.
Therefore, we used only the \ion{Mg}{2} $\lambda$2803 line for the
Voigt profile fitting for both images A and B for this component. The
results are summarized in Table \ref{vpfit}. Due to the lack of
the information on \ion{Mg}{2} $\lambda$2796, the uncertainties are fairly large
especially for the Doppler width. 

For image A, fortunately, we have the information on the Doppler width
with a high precision from the high-resolution optical spectrum (RSB99).
We tried to estimate the Doppler width of \ion{Mg}{2} $\lambda$2803
line from those of \ion{C}{2} (9.8 $\pm$ 0.5 km s$^{-1}$)
and \ion{Si}{2} (6.7 $\pm$ 0.5 km s$^{-1}$) lines. Generally, a
Doppler width $b$ consists of thermal ingredient $b_{\text{th}}$ and
turbulence ingredient $b_{\text{tu}}$:
\begin{equation}
 b^2 = b_{\text{th}}^2 + b_{\text{tu}}^2
\end{equation}
and the thermal ingredient is expressed as :
\begin{equation}
 b_{\text{th}}^2 = \frac{2kT}{m},
\end{equation}
where $k$ is the Boltzman constant, $T$ is the equilibrium temperature of
a gas cloud, and $m$ is the particle mass. From the Doppler widths of
two species of different mass (carbon and silicon), the temperature
and the turbulence Doppler width were estimated as $T=65,000 \pm 18,000$
K, and $b_{\text{tu}} = 2.6\pm2.7$ km s$^{-1}$, respectively, resulting in the
Doppler width of \ion{Mg}{2} of $7.2 \pm 1.3$ km s$^{-1}$. This value is consistent with the Doppler width estimated solely from the \ion{Mg}{2}
$\lambda$2803 line (5$\pm$29 km s$^{-1}$), but with much improved uncertainty. 

\begin{figure}
 \centering
 \includegraphics[width=7cm,clip]{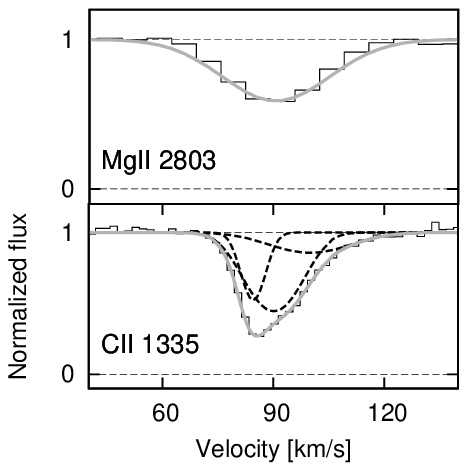}
 \caption{Upper panel: the \ion{Mg}{2} $\lambda$2803 absorption profile of component 3 of image A ($R\sim$10,000). The gray line shows the fitted Voigt profile. Although this velocity component is known to be composed of three sub-velocity components (RSB99), we fitted just one component because the \ion{Mg}{2} absorption profile is convolved with the instrumental function of $\sim$30 kms$^{-1}$ which is much larger than the intrinsic width ($\sim$10 kms$^{-1}$) of the velocity sub-components. Lower panel:the \ion{C}{2} $\lambda$1335 absorption profile of component 3 of image A ($R\sim$70,000). The black line shows the observed normalized spectrum from RSB99. The three black dashed lines show the fitted Voigt profiles while the gray line shows the combined profile.}
 \label{vp3fit}
\end{figure}

\noindent \textit{Component 3.} Because the influence of the telluric
absorption lines is little for component 3 (Figure \ref{telluric}),
which is consistent with the doublet ratio of component 3 for image B as
discussed above, we fitted a Voigt profile to this component for both
\ion{Mg}{2} $\lambda\lambda$2796, 2803 lines simultaneously (Table
\ref{vpfit}). Because the optical absorption profiles are asymmetric for
both images A and C (see Figure \ref{spectra}), RSB99 fitted three
velocity sub-components to the profile of image A. Because they did not
show the fitting results in their paper, we newly fitted the three
sub-components to their data\footnote{The machine-readable data were
kindly provided by Dr. Rauch.} (Figure \ref{vp3fit}) : the fitting
results for the three sub-components are $\log
N$(\ion{C}{2})[cm$^{-2}$]=13.06, 13.49, 12.96, and b=3.5, 9.4, 14.9 km
s$^{-1}$, respectively. The observed \ion{Mg}{2} absorption profile of
component 3 is also found to be slightly asymmetric. Although, at first,
we tried to fit multi-components to component 3, \ion{Mg}{2} absorption
lines with the three components identified by RSB99, the resultant
fitting uncertainties were very large because of the lower spectral
resolution ($\Delta v \sim 30 $ km s$^{-1}$) compared with the widths of
three components ($b < 10$ km s$^{-1}$). Therefore, we treat component 3
as a single component for both images in the fitting with VPFIT.
Although the resultant Doppler width does not have any physical meaning,
the estimated column density is expected to be accurate, at least for
image B, because the \ion{Mg}{2} absorption lines are not saturated for
image B as expected from the doublet ratio.  For image A, though the
\ion{Mg}{2} $\lambda$2803 line is expected to be slightly saturated in
view of the column density (10$^{13}$ cm$^{-2}$) and the expected
smallest Doppler width (9 km s$^{-1}$), the resultant column density of
\ion{Mg}{2} is expected to be reasonably accurate.

\subsection{\ion{Fe}{2} Absorption Lines}
\label{FeIIAbsorptionLines}

Because iron is essential to assess the chemical abundance of the
object, the detection of \ion{Fe}{2} absorption lines for the $z=3.54$
system is important to understand the nature of this high-$z$ gas cloud.
Note that the detection of \ion{Fe}{2} $\lambda$1608 in Keck HIRES
spectra is not reported so far because the past optical observations of
B1422+231 \citep{rau01a,son96} do not cover the wavelength of the
shifted \ion{Fe}{2} $\lambda$1608 absorption line of $z=3.54$ system
($\sim$ 7300 \r{A}). In our Subaru IRCS spectra, two \ion{Fe}{2} lines,
$\lambda$2383, $\lambda$2600, with the largest oscillator strengths were
detected, but only for component 3 of image A, which has a higher column
density for all the $\alpha$-elements.

In order to examine the other velocity components of the \ion{Fe}{2}
absorption lines for each image A and B, we combined spectra for
$\lambda$2344, $\lambda $2383 and $\lambda$2600 lines to improve the S/N
for \ion{Fe}{2} detection. We did not use $\lambda$2587 for this
combining because of the intrinsic weakness and the resultant low
S/N. First we transformed the profiles of $\lambda$2344, $\lambda$2383
lines to that of $\lambda$2600 using the following equation:
\begin{equation}
 s'(\lambda) = 1 - \left( 1 - s ( \lambda \times \frac{\lambda _{2600}}{\lambda _{2344\ \text{or}\ 2383}}) \right) \frac{f_{2600}}{f_{2344\ \text{or}\ 2383}}
\end{equation}
where $s(\lambda)$ is the observed spectrum, $s'(\lambda)$ is the
transformed spectrum, and $f$ is the oscillator strength of the
lines. Before the combining, the spectrum for \ion{Fe}{2} $\lambda$2600
($R=$10,000, $\Delta v = 30\ \text{km s}^{-1}$) is smoothed to match the
spectral resolution to that of $\lambda$2344 and $\lambda$2383
($R=$5,000, $\Delta v = 60\ \text{km s}^{-1}$). Finally each spectrum is
combined with weighting by square of the S/N of the continuum around the
absorption lines.  The resultant spectra are shown at the top of Figure
\ref{Fecomb}. We also plotted the expected absorption lines with gray
lines in Figure \ref{Fecomb} assuming $\log
N$(\ion{Mg}{2})/$N$(\ion{Fe}{2}) = 0.31, which is the value for
component 3 of image A in Table \ref{vpfit} when we assume that the
absorption line is composed of a single velocity component. As a result
of Voigt profile fitting for these detected \ion{Fe}{2} absorption
lines, they are found to have a very large Doppler width ($23\pm 6$ km
s$^{-1}$, Table \ref{vpfit}).

For image A, component 3 is clearly detected again on the combined
spectrum, confirming the detection. A weak absorption line is
newly noticed between the wavelengths of component 1 and component 2. However, because the center velocity has a large offset
from that of the \ion{Mg}{2} line, it is hard to confirm the detection at more than a $3\sigma$ level with the
present data. Similarly, for image B, two dips near the center
velocities of components 1 and 2 are found in the combined spectrum,
 but the detection is tentative because the expected absorption is
much weaker in view of the corresponding weak \ion{Mg}{2} absorption
lines.

For component 3 of image B, compared to the expected absorption profile 
shown with gray line in the right panel of Figure \ref{Fecomb}, the observed combined spectrum is almost
flat and the absorption appears to be significantly less than the
expected amount. The 3$\sigma$ upper limit is calculated as $\log N =
12.1$ for this non-detected component.  From this, the column density
ratio \ion{Mg}{2} to \ion{Fe}{2} for the component 3 of image B is
estimated to be $>0.5$, which appears to be much larger than that of
image A, 0.31. This may suggest a large variance of the column density
ratio, $\log N$(\ion{Mg}{2})/$N$(\ion{Fe}{2}), even in a small scale of
$\sim 8\text{pc}$, which is the projected separation between sightlines
A and B at $z=3.54$.

\begin{figure}
 \centering
 \includegraphics[width=9cm,clip]{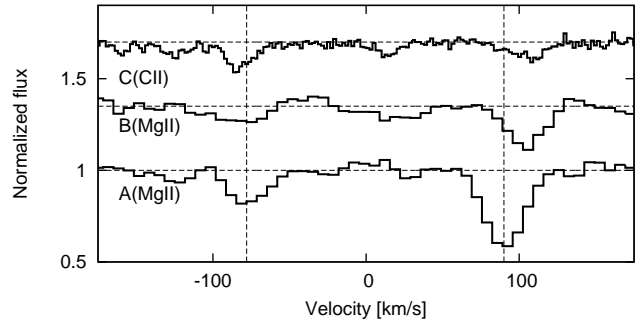}
 \caption{Velocity profiles of the $z=3.54$ system for images $A$(\ion{Mg}{2}, $R=$10,000), $B$(\ion{Mg}{2}, $R=$10,000), and $C$(\ion{C}{2} from RSB99, $R=$70,000) from bottom to top. The velocity is relative to $z=3.53850$. The vertical dashed lines show the peak velocity of components 2 and 3 of image A. The horizontal dashed lines show the normalized continuum level.}%The spectra of image C is smoothed to match the spectral resolution with images A and B.}
 \label{compABC}
\end{figure}

\section{Discussion}

\subsection{Small-scale Structure of Absorbing Gas Cloud at $z=3.54$}

By comparing the results for images A and B, considerable differences
of the column densities of \ion{Mg}{2} absorption lines ($\log (N_A / N_B) \sim$ 0.20$\pm 0.12$ dex and
0.37 $\pm 0.08$ dex for component 2 and component 3, respectively) and
considerable velocity shears 
($|v_A - v_B| \sim$ 7.0$\pm 5.8$ km s$^{-1}$ and 12.3$\pm0.8$ km s$^{-1}$ at $z=3.53850$ for component 2 and
component 3, respectively) are found as shown
in Figure \ref{compABC}.
Considering the alignment of three images A, B, and C, the differences were as expected from RSB99's interpretation of their
optical spectra of images A and C in that the relations of the column
densities and the relative velocities among three images are $\log N_A
> \log N_B > \log N_C$ (assuming $\log 
N(\text{\ion{Mg}{2}})/N(\text{\ion{C}{2}})$ is equal for both images B
and C ) and $|v_A| < |v_B| < |v_C|$ for both components 2 and 3. The projected distances at
$z=3.54$ between images are $d_{AB} = 8.4 h_{70}^{-1}$ pc and $d_{AC}
= 22.2 h_{70}^{-1}$ pc. 
This very high spatial resolution (10 pc at
$z=3.54$ corresponds to $\sim$ 1 mas angular resolution for direct
imaging) shows the power of the gravitational lensing (RSB99).

B1422+231 also has a QSO absorption system at slightly higher redshift (z=3.624) and the transverse distance reaches 1 pc, which is the smallest separation for QSO absorption systems ever studied with gravitationally lensed QSOs \citep{jil95, rau01a}, though the system shows few variations of absorption lines among multiple images \citep[see Figure 5 of][]{rau01a} and this system is likely to be associated with the QSO itself. Therefore, the $\sim$ 10 pc structure for the $z=3.54$ system is the smallest spatial structure ever detected for QSO absorption systems.

\citet{ell04} observed three lensed images of gravitationally lensed QSO APM08279+5255(z=3.911) with \textit{Hubble Space Telescope (HST)}. They detected many metal absorption lines at $1.1 < z < 3.8$, which correspond to the transverse distance, from 30  $h_{70}^{-1}$ pc to 2.7 $h_{70}^{-1}$ kpc. They found large variations of EWs for lower ionization systems, which are traced with \ion{Mg}{2} doublet lines, even on the spatial scale of a few hundred pc while the higher ionization systems, which are traced with \ion{C}{4} doublet lines, show less variations of EWs \citep{rau01a,ell04}. Therefore, low ionization systems should reflect small scale gas clouds, which are likely to be related to star formation activities in galaxies. Because this $z=3.54$ system's spatial scale ($\sim$ 10 pc) corresponds to that of the typical Galactic gas clouds, such as giant molecular clouds, \ion{H}{2} regions, and SNRs \citep{rau99}, the $z=3.54$ system toward B1422+231 is a very precious target that enables the study of such Galactic-scale objects in detail at the galaxy-forming epoch.

\begin{figure*}
 \centering
 \includegraphics[width=6cm,clip]{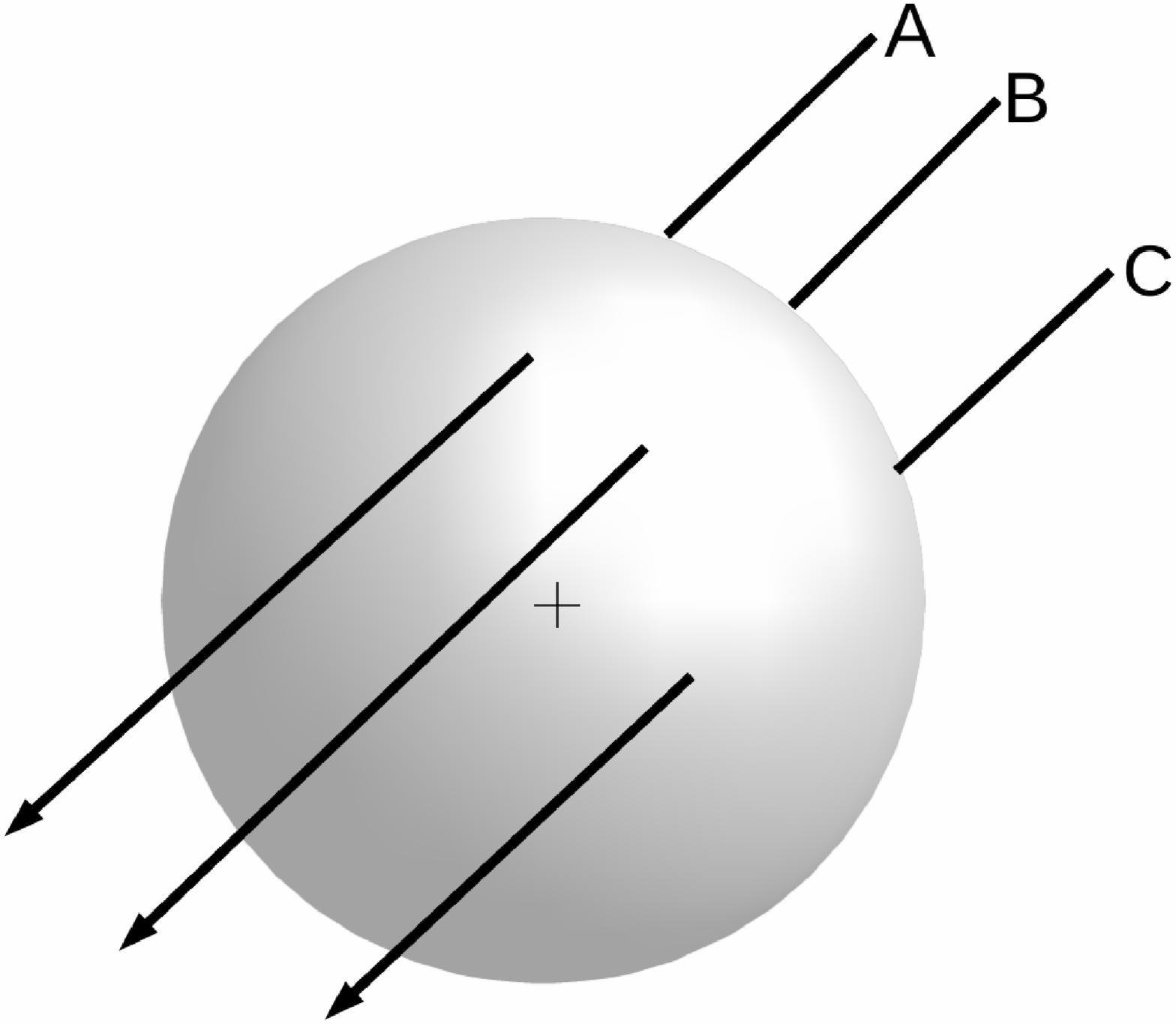}
 \includegraphics[width=5cm,clip]{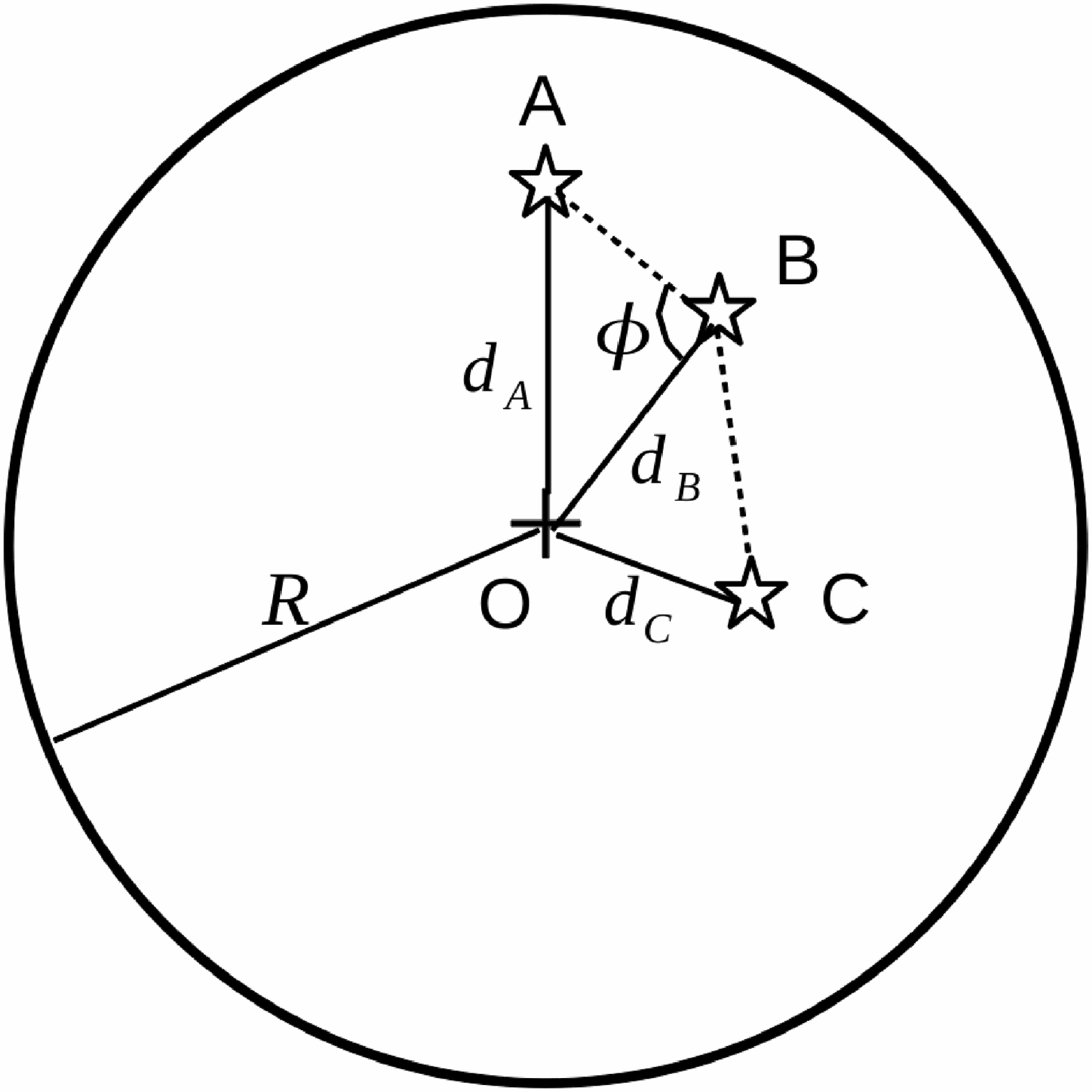}
 \includegraphics[width=6cm,clip]{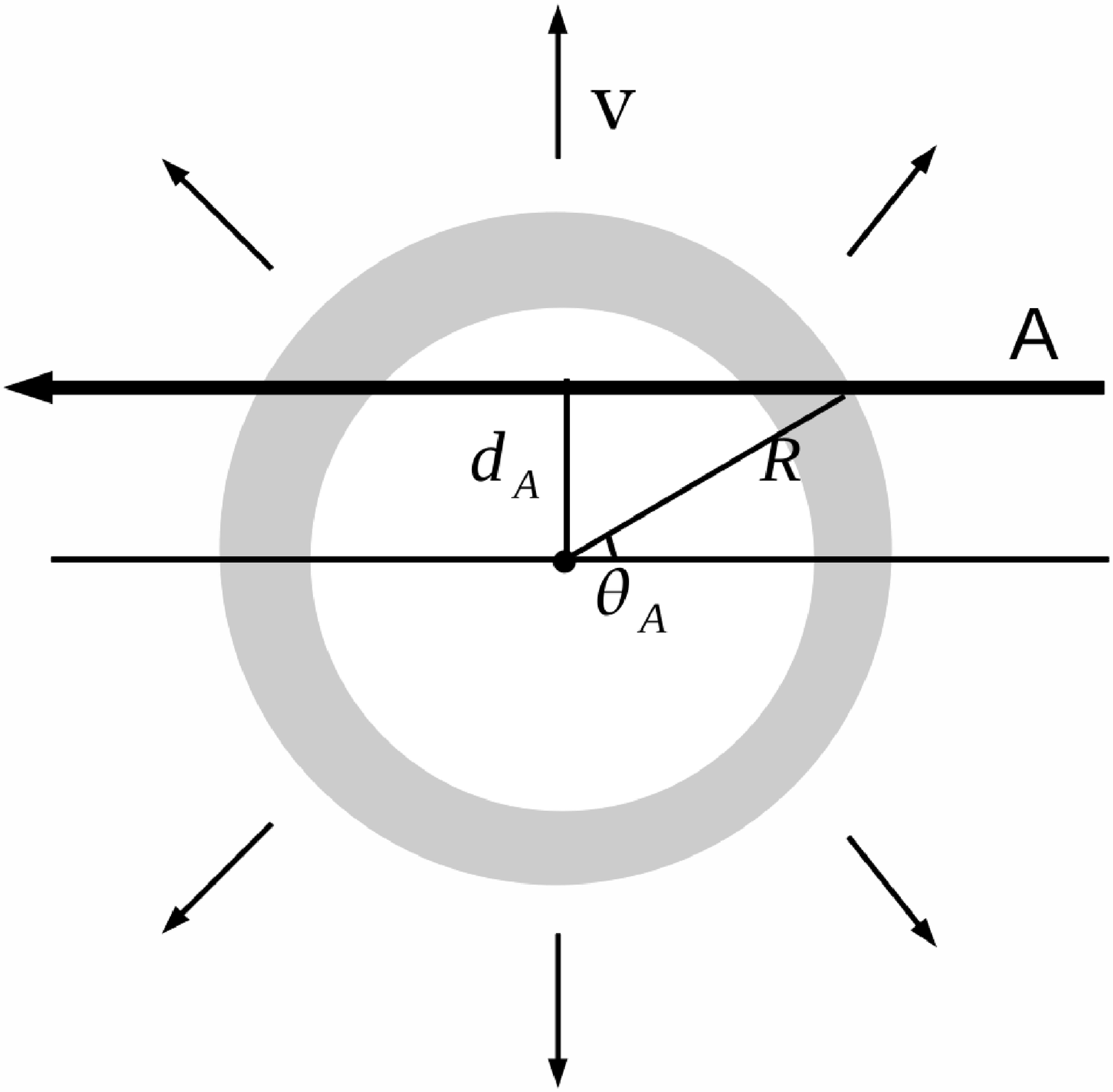}
 \caption{Images of the expanding shell to illustrate the formulation in the text (Section 5.2.1) for the case where information from three sightlines, A, B, and C, is available. Left panel: the 3D image of sightlines that pass through the shell. Middle panel: the expanding shell seen from the direction of sightlines, showing the definitions of $d_A, d_B, d_C$, and $ \phi$. Right panel: the cross-section of the shell that contains the sightline of image A and the center of the shell, showing the definition of $\theta _A$. Similarly, $\theta _B$ and $\theta _C$ are defined for the sightlines of images B and C, respectively.}
 \label{shell3d}
\end{figure*}

\begin{figure*}
 \centering
 \includegraphics[width=6cm,clip]{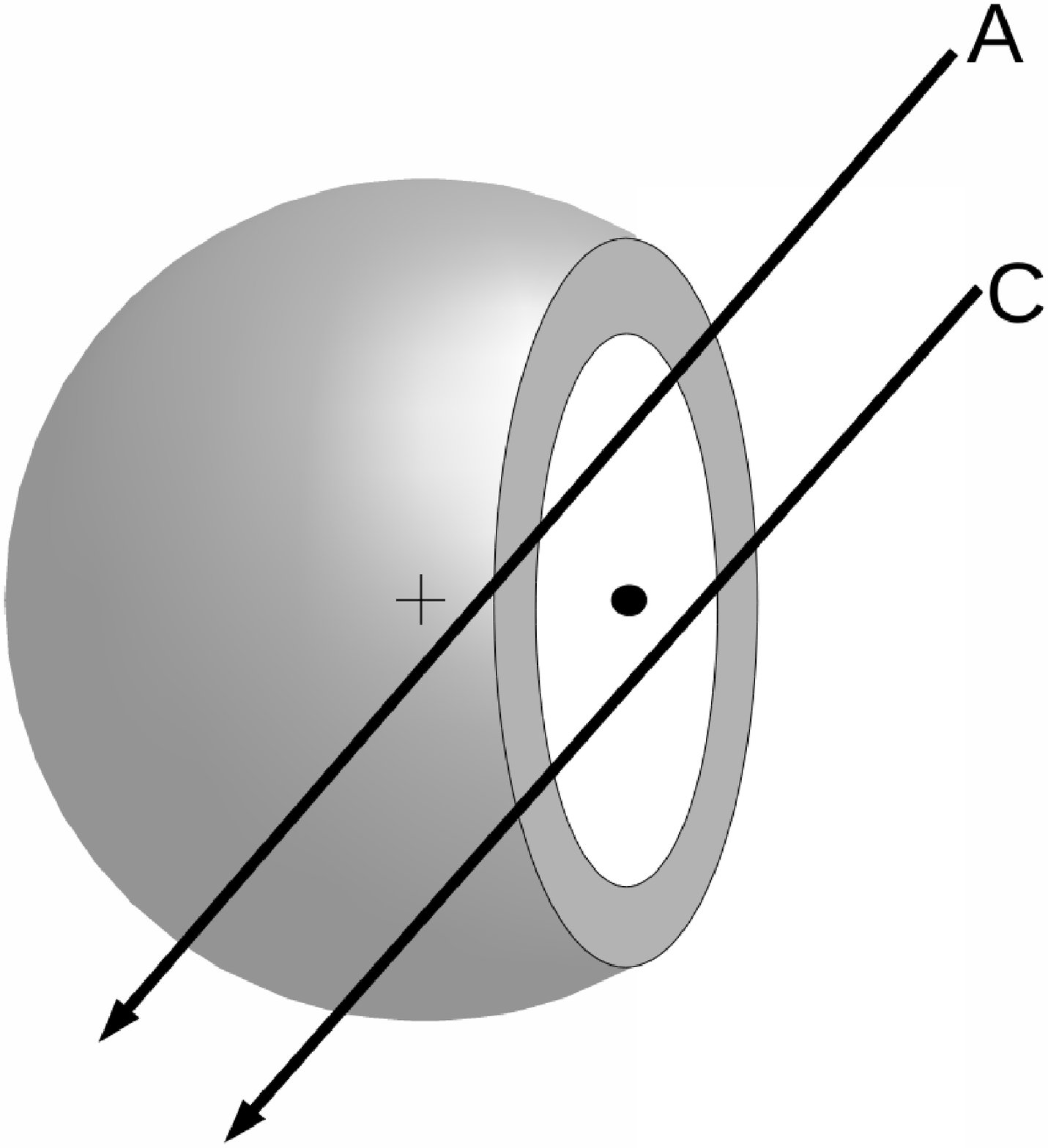}
 \includegraphics[width=5cm,clip]{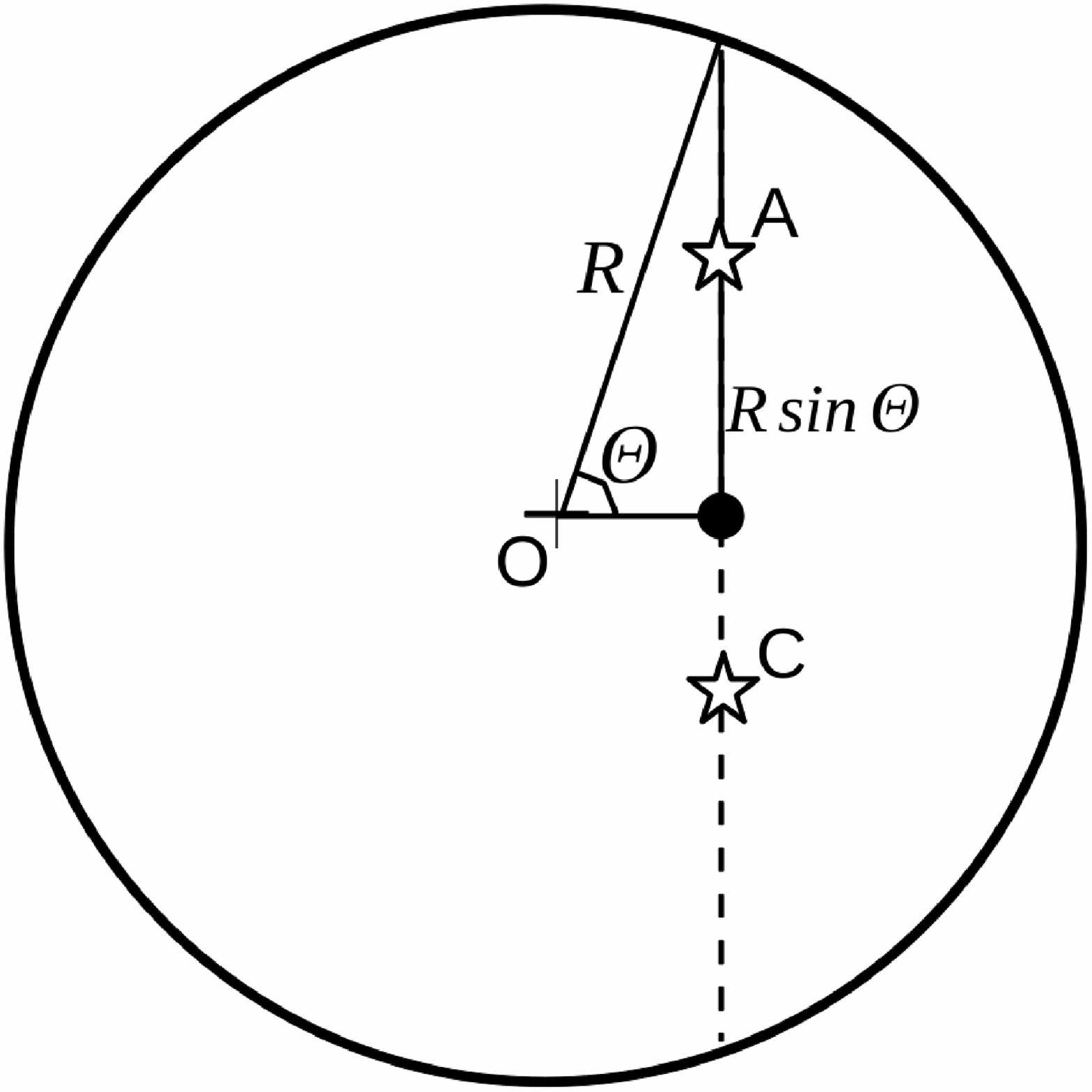}
 \includegraphics[width=6cm,clip]{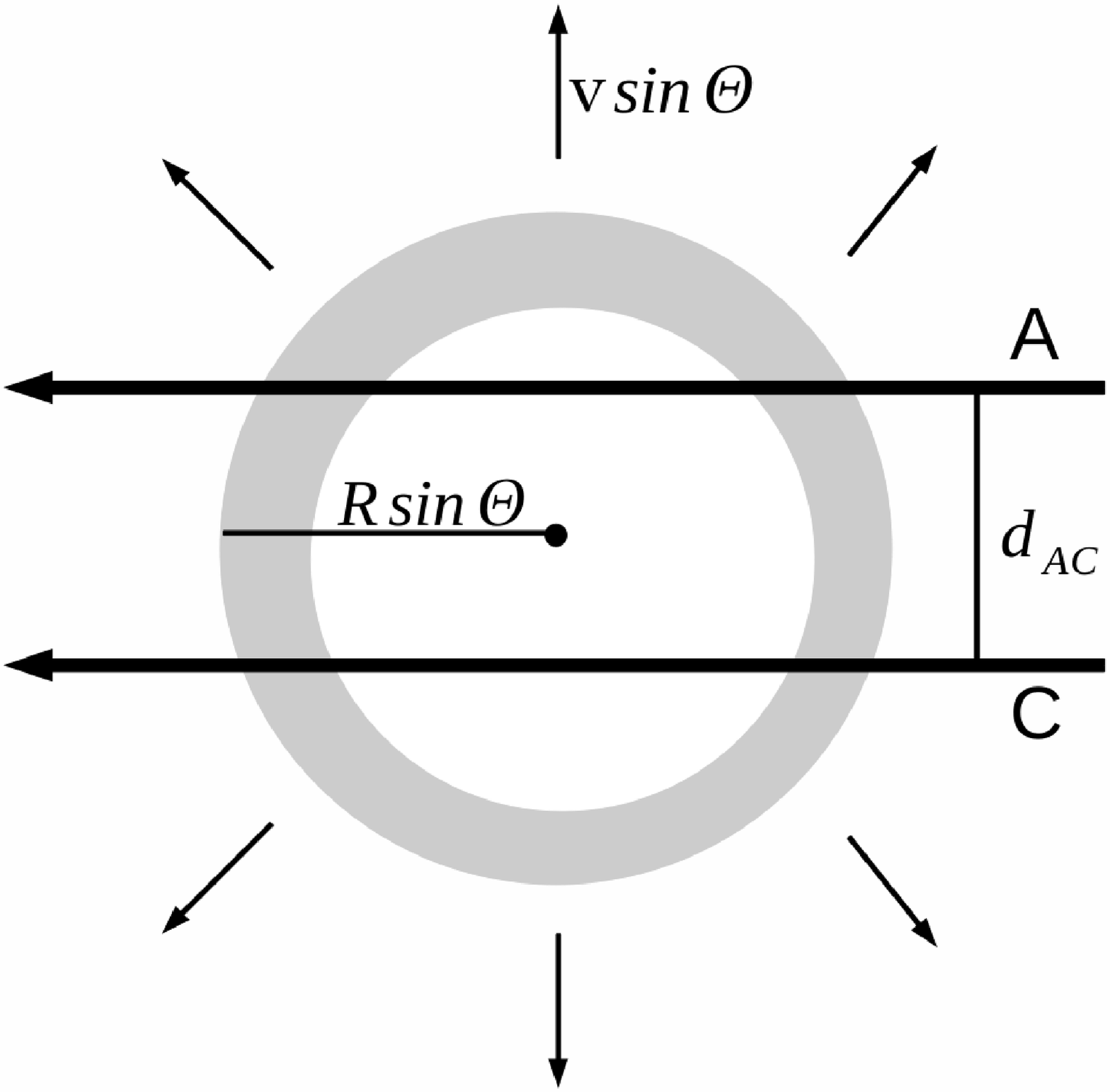}
\caption{Images of the expanding shell to illustrate the formulation in the text (Section 5.2.1) for the case where information from two sightlines, A and C, is available. Left panel: the 3D image of sightlines that pass through a circle that is off the center of the shell. Middle panel: the expanding shell seen from the direction of sightlines, showing the definition of $\Theta$. The radius of the cross-section is $R\sin \Theta$. Right panel: the cross-section of the shell that contains the sightlines of images A and C. Though we can constrain $R\sin \Theta$ and $v\sin \Theta$, we cannot infer the radius $R$ and expansion velocity $v$ due to the lack of information of another independent sightline, such as the sightline B.}
 \label{shell2d}
\end{figure*}

\subsection{Expanding SNR Shell}

\subsubsection{Shell Model}

The column densities and the velocities of \ion{Mg}{2} absorption
lines of components 2 and 3 vary among three sightlines as $\log N_A >
\log N_B > \log N_C$ and $|v_A| < |v_B| < |v_C|$. What types of object
can generate these \textit{systematic} variances on such a small scale? Based
on the nearly symmetric velocity profiles of components 2 and 3 with
respect to the systemic velocity that corresponds to $z=3.53850$ that are seen in optical spectra of both images A and B (see the profiles for
\ion{C}{2} and \ion{Si}{2} in Figure \ref{spectra}), RSB99 suggested
that the components 2 and 3 of the $z=3.54$ system can be an expanding
shell, such as an SNR. When the sightline passes through the center of a gas
cloud of the expanding shell, the observed velocity becomes highest
because the gas expands along the direction of the sightline while the
column density becomes lowest because the path of the sightline in the
shell is shortest at the center. On the other hand, when the sightline
passes through the outer edge, the observed velocity becomes lowest, while the
column density becomes highest. Applying this model to the $z=3.54$
system, RSB99 found that the sightlines of images A and C pass near the
outer edge and near the center of an expanding gaseous shell,
respectively. The absorption lines of image B, which is newly observed
in this study, shows intermediate values for both column density and
relative velocity; this systematic kinematics seen in this gas cloud
support the idea of RSB99 that this $z=3.54$ system is truly an
expanding shell. Although a contracting (or collapsing) shell is an
alternative choice, such an astronomical object is unlikely because the
central object of the shell has to pull all the gas symmetrically in
a subtle manner: normally such gas should fall through a disk or
infalling envelope (not a shell). In fact, no such object is known in
the Galaxy and near-by galaxies (see the listed examples of
astronomical objects in Section 5.2 in \citealt{rau02}). 

Following RSB99 and \citet{rau02}, we will consider a model of a
three-dimensional (3D) expanding shell (Figure \ref{shell3d}) that has a
radius of $R$ and an expansion velocity of $v_{\text{exp}}$. Those parameters
can be constrained by two kinds of observables: the physical distance
between two sightlines at the redshift (e.g., $d_{AB}$) and the
velocity difference of the two absorption components seen in a single
sightline (e.g., $\Delta v_A$). In case only two sightlines are
available, only the parameters for the expansion ``ring'' that is on the
plane of the two sightlines can be constrained: one is the
radius of the ring ($R\sin\Theta$) and the other is the expansion
velocity of the ring ($v_{\text{exp}}\sin\Theta$), where $\Theta$ is the
declination of the ring on the sphere (Figure \ref{shell2d}). Although
the relation between $R\sin\Theta$ and $v_{\text{exp}}\sin\Theta$ can be
obtained, the 3D radius ($R$) and the expansion
velocity ($v_{\text{exp}}$) can never be determined because of the complete
lack of the information on the absolute location of the ring
($\Theta$) on the sphere: only the possible range of $R\sin\Theta$ and
the lower-limit of $v_{\text{exp}}\sin\Theta$ can be obtained (RSB99).

On the other hand, in case {\it three} independent sightlines (not on
a single plane) are available, we can put a strong constraint on the
expanding sphere. Since the absolute location of the planes that
contain the sightlines cannot be determined in a unique way, $R$ and
$v_{\text{exp}}$ still cannot be determined. {\it However}, if the $R$ value
is fixed, the other parameter $v_{\text{exp}}$ is determined because two
$\Theta$s for the sets of two sightlines (e.g., A/B and B/C) can be
determined from the two equations for two rings. Therefore, the
relation between $R$ and $v_{\text{exp}}$, which is useful to elucidate the
astronomical nature of the shell, can be obtained as a final product
(see the detailed description in Appendix of \citealt{rau02}). In
fact, \citet{rau02} suggested that the $z=0.5656$ absorption system in
the three sightlines of gravitationally lensed QSO Q2237+0305 is also
an expanding shell because the absorption lines showed
variances similar to those of the $z=3.54$ system. They analysed the relation between
the radius and expansion velocity of the 3D shell and
concluded that the expanding shell at $z=0.5656$ can be interpreted as
a supershell or a superbubble of 1-2 kpc size that is frequently
observed in the Galaxy and extra-galaxies.

Based on the radial velocities in the two lines of sight (images A and
C), RSB99 managed to constrain the parameters in a ring for the
$z=3.54$ system as 9 $h_{70}^{-1} < R\sin\Theta <$ 34 $h_{70}^{-1}$ pc and $v_{\text{exp}}\sin\Theta
> 98\ \text{km s}^{-1}$. Now, with the additional information of image B, we
can obtain the relation of $R$ and $v_{\text{exp}}$ of the expanding shell
at $z=3.54$. Note that the model parameters can be completely
determined if we have four independent sightlines. Because B1422+231
has four independent gravitationally lensed images, future
observations of the fourth image would be very valuable.

We formulated the geometry of the expanding shell as shown in Figure
\ref{shell3d} to obtain the radius-velocity relation of the $z=3.54$
system. The difference of velocities of components 2 and 3 in the
sightline of image A, $(\Delta v)_A$, and the distance from the center
of the shell to the sightline A, $d_A$, can be expressed as
\begin{gather}
 (\Delta v)_A = 2 v_{\text{exp}} \cos \theta _A \\
 d_A = R \sin \theta _A ,
\end{gather}
where $\theta _A$ is the angle from the center of the shell to the
sightline A (see Figure \ref{shell3d}). The equations for images B and C
can be given similarly, resulting in six equations in total. Next, we
can derive two equations about the geometry of triangle ABC :
\begin{gather}
 d_A^2 = d_B^2 + \overline{AB}^2 + 2d_B \overline{AB} \cos{\phi} \\
 d_C^2 = d_B^2 + \overline{BC}^2 + 2d_B \overline{BC} \cos{[\angle ABC - \phi]} ,
\end{gather}
where $\overline{AB}$ and $\overline{BC}$ are projected distances
between images at $z=3.54$, $\phi$ is the angle between BA and BO, and
$\angle ABC$ is the angle between BA and BC. Now, the parameters
$(\Delta v)_A$, $(\Delta v)_B$, $(\Delta v)_C$, $\overline{AB}$,
$\overline{BC}$, $\angle ABC$, are observed and the nine parameters $R$,
$v_{\text{exp}}$, $d_A$, $d_B$, $d_C$, $\theta _A$, $\theta _B$, $\theta _C$,
$\phi$ are not determined. If these eight equations are combined, the
$R(v_{\text{exp}})$ relation can be obtained as a solution\footnote{The analytic
solution is given in Equation (A5) in \citet{rau02}.}. 

\begin{figure}[t]
 \centering
 \includegraphics[width=9cm,clip]{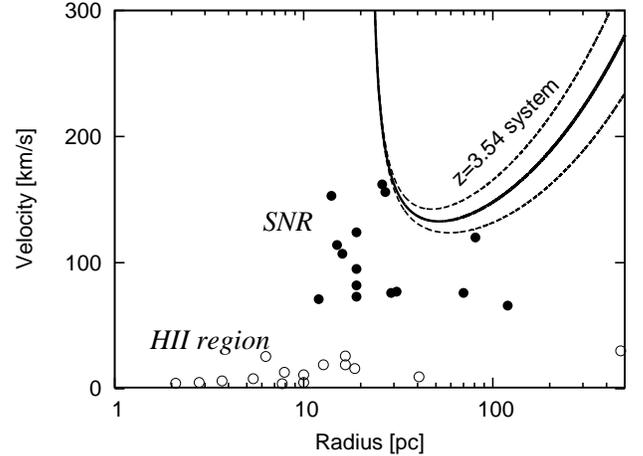}
 \caption{Relation of radius and expansion velocity of the expanding shell model. The solid line shows the relation for the $z=3.54$ system in the case of $(\Delta v)_B=188$ km s$^{-1}$, which is the most likely value from the observation (see Section 5.2.1 for detail). The dotted lines along the solid line are the case of $(\Delta v )_B=185,191$ km s$^{-1}$, showing 1$\sigma$ uncertainty of our calculation. The filled and open circles show the radius and expansion velocity of the observed SNRs and \ion{H}{2} regions in our Galaxy \citep{koo91,shi90,kot02,arn07,dai07,cap08,cic08}.}
 \label{rad-vel}
\end{figure}

For the $z=3.54$ system, the projected distances are $\overline{AB} =
8.4 h_{70}^{-1} $ pc, $\overline{BC} = 14.4 h_{70}^{-1} $ pc, and
$\overline{AC} = 22.2 h_{70}^{-1} $ pc at $z=3.54$. The angle between AB
and BC is 153$^\circ$. The velocity differences between components 2 and
3 are $(\Delta v)_A = 168\pm1$ km s$^{-1}$, $(\Delta v)_B = 188 \pm 3$
km s$^{-1}$, and $(\Delta v)_C = 197\pm1$ km s$^{-1}$. By putting those
values into the equations, we obtained the function $R(v_{\text{exp}})$
that is shown with a solid line in Figure \ref{rad-vel} along with two
dashed lines that show the 1$\sigma$ uncertainty range of our
calculation : each line corresponds to the cases of $(\Delta v)_B =$ 185
km s$^{-1}$, 191 km s$^{-1}$. Here, we are only concerned with the
uncertainty of $(\Delta v)_B$ because it is the largest among six
observables : the available spectrum of image B is only our
lower-resolution near-infrared data while the available spectra of the
other two images A and C are higher-resolution optical data of RSB99.

\subsubsection{The SNR Origin of the Shell}

Figure \ref{rad-vel} shows that the radius of the shell is larger than
$\sim20$ pc and the expansion velocity is larger than $\sim120$ km
s$^{-1}$. Although there is no a priori constraints, the shortest radius
($R \le 30$ pc) is unlikely because it requires: (1) an exact
alignment of the shell with the three sightlines that have similar
separations as the shell diameter and (2) very high velocity ($v_{\text{exp}} >200$ km s$^{-1}$) that appears to be too fast for an \ion{H}{1}
shell. The largest radius ($R \gg 100$ pc) is also unlikely since it
requires similarly very high expansion velocity and all three sightlines must pass near the edge of expanding shell, exactly. Therefore, the bottom
of the curve ($R \sim50-100$ pc, $v_{\text{exp}}\sim130$ km s$^{-1}$) would be
the natural choice of the radius and the expansion velocity of the shell (see
the original discussion in \citealt{rau02} for another QSO absorption system).

To compare to the known astronomical shell-like objects, we also
plotted the radii and expansion velocities observed for the Galactic
\ion{H}{2} regions (\citealt{shi90,kot02,arn07,dai07,cap08,cic08}) and
SNRs that have a structure of expanding shell: the data points of SNR
are from \citet{koo91}, who observed 15 old SNRs with \ion{H}{1} 21cm
line in the disk of our Galaxy. As a result, the location of the $z=3.54$ system on
the plot is found to be quite consistent with SNRs while all the \ion{H}{2}
regions show much smaller expansion velocity less than 20 km s$^{-1}$.
In fact, the distributions of SNRs match well with the bottom of the
curve for the $z=3.54$ system, which is quite consistent with the
above consideration that the bottom of the curve is the most likely
location of the shell.
Therefore, we concluded that the shell is truly an SNR shell,
confirming RSB99's initial suggestion in more solid way. 

\subsubsection{Physical Parameters of the SNR Shell}

Assuming $R=50-100$ pc and $v_{\text{exp}}=130$ km s$^{-1}$ as the radius and the 
expansion velocity of the observed shell, we compare the observed SNR
shell with the typical self-similar expansion model to estimate the main
physical parameters of the shell. Then, we compare the estimated
physical parameters to those of typical SNRs to check the consistency
of the SNR interpretation. In the following, we will utilize the
formulation described in text-books by \citet{tie05} and
\citet{dra11} to first estimate the age ($t$) of the SNR shell, and
finally the total energy ($E$) of the shell, along with an independent
estimate of mass ($M$) of the shell
from the column densities of the absorption lines.

If the self-similar concept is introduced with a simple
power laws ($R \propto t^{\eta}$), 
the age of the shell can be roughly estimated from $R$ and $v_{\text{exp}}$ as $t\sim 10^5 $yr for $z=3.54$ system since $t\sim R/v_{\text{exp}}$. The SNR shell of this age is in the radiative expansion phase (snowplow phase) rather than adiabatic expansion phase (Sedov phase). In this case, 
$\eta = 2/7$ and the age
$t$ can be derived with
\begin{equation}
 t = 1.1 \times 10^5 \text{yr} \left( \frac{R}{50\text{pc}}\right) \left( \frac{v_{\text{exp}}}{130 \text{km s}^{-1}} \right) ^{-1} . \label{age}
\end{equation}
For the observed $v_{\text{exp}}$ (130 km s$^{-1}$) and the range of $R$ (50 - 100
pc), the age of the $z=3.54$ shell is estimated as:
\begin{equation}
1.1 \times 10^5 \text{yr} \leq t \leq 2.2 \times 10^5 \text{yr} . 
\end{equation}
In this snowplow phase, the
radius and expansion velocity are modeled as
\begin{equation}
 R(t) = 21.6 \text{pc} \left( \frac{E_0}{10^{51} \text{erg}} \right)^{\frac{11}{45}} \left( \frac{n_0}{1\text{cm}^{-3}}\right) ^{-\frac{11}{45}} \left( \frac{v_{\text{exp}}(t)}{250\,\text{km s$^{-1}$}} \right) ^{-\frac{2}{5}} ,\label{radius}
\end{equation}
where $E_0$ is the total energy of a supernova, $n_0$ is the number density of the interstellar gas {\it before the supernova explosion}.

Next, we will discuss the mass of the shell ($M_{\text{tot}}$). 
RSB99 estimated the number density ($n$) and the size
along the line of sight ($L$) for the component 3 of image A using
photoionization model as
\begin{gather}
 0.16 \text{ cm}^{-3} \leq n \leq 1.6 \text{ cm}^{-3} \label{numberR} \\
 0.015 \text{ pc} \leq L \leq 1.6 \text{ pc} . \label{thickR}
\end{gather}
From these parameters, RSB99 estimated the mass of a gas cloud ($M_{\text{tot}}$) assuming two geometrical cases ; first is a homogeneous cylindrical slab with a thickness $L$ and radius $\overline{AC}$ as a lower limit of the mass; second is a spherical cloud with a radius $\overline{AC}$ as an upper limit of the mass:
\begin{equation}
 0.4M_\odot \leq M_{\text{tot}} \leq 2700M_\odot . \label{rsbmass}
\end{equation}
Here, we newly obtained an additional constraint on the radius $R$ of the gas cloud as $R \sim 50 - 100$ pc. Assuming that the gas cloud has a shape of spherical shell with an average radius of $R$, thickness of $L$, and number density of $n$, we can calculate the total mass $M_{\text{tot}}$ of the shell with 
\begin{equation}
M_{\text{tot}} = 4 \pi R^2 L n m_{\text{\ion{H}{1}}} \mu ,
\end{equation}
where $m_\text{\ion{H}{1}}$ is the mass of a hydrogen atom particle ($m_{\text{\ion{H}{1}}} = 1.66\times 10^{-24} $g) and $\mu$ is the reduced mass ($\mu = 4/3$). Here we ignored the effect of $\cos \theta _A$ on the thickness of the shell (see Figure \ref{shell3d}) in view of the large uncertainty of $L$. Using the value range of $n$ and $L$ (Equation \ref{numberR} and \ref{thickR}) and our result on the radius ($50 <R< 100\ \text{pc}$), $M_{\text{tot}}$ is estimated as:   
\begin{equation}
25 M_\odot \leq M_{\text{tot}} \leq 1050 M_\odot .
\end{equation}
This is consistent with the original mass estimate by RSB99 (Equation \ref{rsbmass}), but narrows the mass range by two orders of magnitude.

Because the uncertainty of this estimate is quite large, we try an alternative mass estimate in the following.
We will estimate the mass in two ways here, using the column density of \ion{H}{1} or \ion{Mg}{2}, in order to check the consistency. In both estimates, we will use the column density of component 3 of image A because the ionization parameter, $\log U$, of this component was specifically estimated as $\log U = -4.4$ (RSB99) which is low enough that we can estimate the mass simply from observed column density without any ionization correction. Although the column density of this component may not be the representative value of the shell, we would not expect a large uncertainty of more than an order of magnitude in view of the column density variation among the components seen in images A, B and C. 

First, the \ion{H}{1} mass of the shell can be simply calculated as :
\begin{equation}
  M_{\text{\ion{H}{1}}} = 4 \pi R^2 N_{\text{\ion{H}{1}}} m_{\text{\ion{H}{1}}} \label{himass}
\end{equation}
where $R$ is the size of the shell and $N_{\text{\ion{H}{1}}}$ is the observed \ion{H}{1} column
density.
Using $\log N$(\ion{H}{1})=$16.05 \pm 0.15$ for component 3 of image A (RSB99), the \ion{H}{1} mass is calculated as
\begin{equation}
 2.8 M_\odot \leq M_{\text{\ion{H}{1}}} \leq 11 M_\odot \label{mass_hi}
\end{equation}
Because this gas cloud is optically thin, almost all of the hydrogen is likely to be ionized. To estimate the total mass of the shell from the \ion{H}{1} mass, we must evaluate the degree of ionization. In \citet{don91}, the fraction of \ion{H}{1} is calculated with
\begin{equation}
 \frac{n(\text{\ion{H}{1}})}{n_H} = (4.6\times 10^{-6}) U ^{-1.026}, -4.7 < \log U < -1.8,
\end{equation}
where $n$(\ion{H}{1}) is the number density of only \ion{H}{1} and $n_\text{H}$ is the total number density of hydrogen atom that includes \ion{H}{2}. Since the ionization parameter of the component 3 of image A is estimated as $\log U = -4.4$ (RSB99), $n(\text{\ion{H}{1}}) / n(\text{H})$ is calculated as 0.15. Then, the total hydrogen mass is estimated as:
\begin{equation}
 19 M_\odot \leq M_H \leq 75 M_\odot .
\end{equation} 
Finally, the total mass of the shell can be calculated by multiplying the reduced mass, $\mu$, as:
\begin{equation}
 25 M_\odot \leq M_{\text{tot}} \leq 99 M_\odot. \label{totmasshi} 
\end{equation}
This range is consistent with the typical scrambled gas mass of the observed SNR (10 - 1000 $M_\odot$), with a radius from about 10 to a few 100 pc \citep{koo91}. 

Next, we attempt one more independent estimate of the total mass of the shell based on the column density of \ion{Mg}{2} instead of \ion{H}{1}. The total \ion{Mg}{2} mass in the shell can be estimated using Equation (\ref{himass}) but after replacing \ion{H}{1} to \ion{Mg}{2}. For column density $N_{\text{\ion{Mg}{2}}}$, we use the value of component 3 of image A because this component is examined in detail by RSB99. As a result, the total \ion{Mg}{2} mass is calculated as
\begin{equation}
 0.06 M_\odot < M_{\text{\ion{Mg}{2}}} < 0.24 M_\odot \label{massofMgII}
\end{equation}
for the assumed range of radius. 

The total mass of the shell can be estimated first assuming that all the
magnesium is in the form of \ion{Mg}{2}. This assumption is reasonable because the
\ion{Mg}{1} absorption lines are not detected (the 3$\sigma$ upper
limit is calculated as $\log N (\text{\ion{Mg}{1}}) [\text{cm}^{-2}]< 11.1$; see Section 4.1 or Table \ref{vpfit}. Although the possibility of the existence of a significant amount of \ion{Mg}{3} cannot be dismissed, we ignored the higher ionization states because the ionization parameter $\log U$ of this component is estimated to be quite low as $\log U = -4.4$ (RSB99) based on the photoionization modeling of \citet{don91}. We assumed the solar
abundance, which is suggested by RSB99 for component 3 based on
the photoionization modeling by \citet{don91}. 
With the solar abundance of magnesium \citep[0.13\% in mass; ][]{gre10}, 
the total mass $M_{\text{tot}}$ is estimated from Equation (\ref{massofMgII}) as
\begin{equation}
 47 M_\odot < M_{\text{tot}} < 188 M_\odot. \label{mass_tot}
\end{equation}
This mass range is pretty much consistent with the estimate from \ion{H}{1} column density, Equation (\ref{totmasshi}).

From the estimated radius and expansion velocity, we can finally constrain the energy of supernova explosion using Equation (\ref{radius}). The remaining parameter in this equation, $n_0$, which is the number density of interstellar medium around the supernova before explosion, can be estimated assuming that the shell consists of all of the gas that existed in the sphere with radius $R$ before explosion, with the following equation:
\begin{equation}
 n_0 = \frac{3M_{\text{tot}}}{4 \pi R^3 m_H \mu}.
\end{equation}
With the range of $50 \leq R \leq 100 \text{ pc}$ and $M_{\text{tot}}$ from Equation (\ref{mass_tot}), $n_0$ is calculated as:
\begin{equation}
 1.8 \times 10^{-3} \text{cm}^{-3} < n_0 < 3.6 \times 10^{-3} \text{cm}^{-3}
\end{equation}
Then, the energy of supernova explosion, $E_0$, is calculated with Equation (\ref{radius}): 
\begin{equation}
 3.8 \times 10^{49} \text{erg} < E_0 < 3.2 \times 10^{50} \text{erg} \label{energySNR}
\end{equation}
The estimated energy is roughly consistent with the energy of supernova explosion, $\sim 10^{51}$ erg \citep{tie05,dra11}. The slight difference can be attributed to the left-over gas inside the shell \citep[see e.g.,][]{koo91} that can effectively increase $n_0$, thus $E_0$ through Equation (\ref{radius}). 

In this subsection, we have gone through the physical parameters
 of the $z=3.54$ system as an expanding shell of an SNR. The expanding shell model and estimated physical parameters
appear to be quite consistent with the properties of SNRs observed in
the galaxy. With our calculation, this system is likely to be an SNR of about 0.1 Myr
with a radius of 50 $-$ 100 pc, an expansion velocity of about 130 km
s$^{-1}$, and the total energy of $10^{50}$
erg. Therefore, we conclude that this $z=3.54$ system is truly an SNR.

\begin{figure}
 \centering
 \includegraphics[width=8cm,clip]{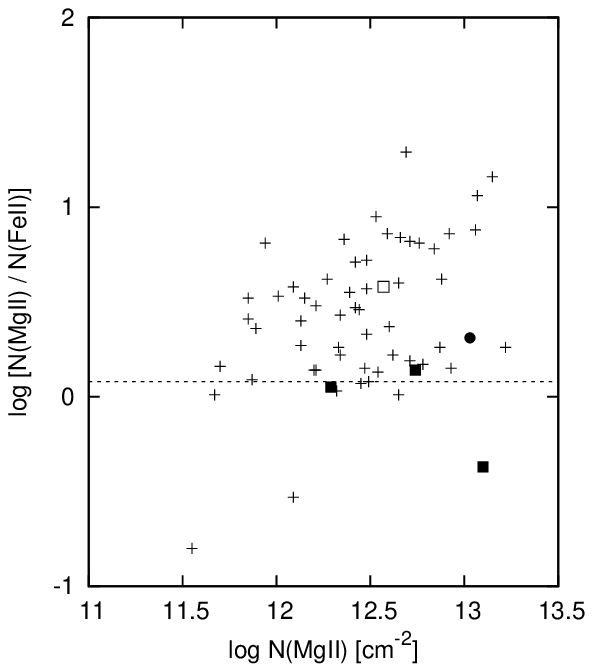}
 \caption{$N$(\ion{Mg}{2}) vs. $N$(\ion{Mg}{2})/$N$(\ion{Fe}{2}) of weak \ion{Mg}{2} systems. The crosses are from \citet{nar08}, the squares are from \citet{rig02}, and a filled circle shows the component 3 of image A of the $z=3.54$ system. The filled squares show confirmed Fe-rich systems while a open square shows confirmed non Fe-rich system \citep{rig02}. The dotted line shows the solar abundance ratio \citep{asp05}. The $z=3.54$ system (filled circle) is located in a horizontal branch near the solar value, where the confirmed Fe-rich systems (filled squares) are distributed.}
 \label{narayanan}
\end{figure}

\subsection{Type Ia Supernova ?}

Recall the very broad profile of \ion{Fe}{2} absorption lines of
component 3 in image A, which is described in Section 4.2. What
does this feature mean in the SNR interpretation of the $z=3.54$ system?
To answer this question, we first discuss the iron richness of the SNR shell, then examine the broad absorption feature in a more rigorous
way to suggest that the iron is localized in the SNR shell. Finally, we conclude that the SNR shell is related to an SN Ia.

\subsubsection{Fe Richness}

The amount of iron in the gas cloud is crucial to investigate the
chemical enrichment by supernovae.  Figure \ref{narayanan} shows the
distribution of $\log N(\text{\ion{Mg}{2}}) / N(\text{\ion{Fe}{2}})$
versus $\log N$(\ion{Mg}{2}) for weak \ion{Mg}{2} systems. The crosses
are from \citet{nar08}, who studied 100 weak \ion{Mg}{2} systems at
$0.4<z<2.4$ using VLT data. The four squares are weak \ion{Mg}{2}
systems studied with Keck data by \citet{rig02}. Note that neither
sample includes the systems whose \ion{Fe}{2} are not detected. The
filled and open squares are confirmed Fe-rich and non-Fe-rich systems
based on their photoionization modeling using CLOUDY, respectively. The
filled circle shows the $z=3.54$ system ($\log N(\text{\ion{Mg}{2}}) /
N(\text{\ion{Fe}{2}})$=0.31$\pm$0.07). The dotted line shows the solar
abundance ratio, $\log [N(\text{Mg}) / N(\text{Fe})]_\odot$ = 0.08
\citep{asp05} for the case that all Mg and Fe atoms are in the
\ion{Mg}{2} and \ion{Fe}{2} ionization states.  The value of the
$z=3.54$ system is found to be closer to the solar value than those for
the other weak \ion{Mg}{2} systems with similar
$N$(\ion{Mg}{2}). \citet{nar08} suggested that any system near the solar
value (dashed line in Figure \ref{narayanan}) are truly Fe-rich systems.

\citet{rig02} suggested that their three Fe-rich systems ($N$(\ion{Fe}{2})$\sim$$N$(\ion{Mg}{2}))
have small sizes of $\sim 10$pc, and high metallicity of $> 0.1 Z_\odot$ (see filled squares in Figure \ref{narayanan}). They specifically predicted that the $z=3.54$ system toward B1422+231
should show strong \ion{Fe}{2} lines ($N$(\ion{Fe}{2}) $\sim$
$N$(\ion{Mg}{2})) in view of the small spatial structure ($\sim$10 pc)
and high metallicity inferred by RSB99. In fact, $\log N(\text{\ion{Mg}{2}}) / N(\text{\ion{Fe}{2}})$ of the $z=3.54$ system is found to be similarly Fe-rich as the three Fe-rich systems (see Figure \ref{narayanan}), suggesting solar to sub-solar metallicity.

All those results support the
iron-richness of the SNR shell at $z=3.54$.
The high iron abundance of the SNR naturally suggests that it is an
SN Ia. In fact, \citet{rig02} suggested that the three iron-rich
systems in their samples have [$\alpha$/Fe]$<0$ and have been enriched by SNe Ia because
high iron column density that is similar to magnesium column density
cannot be explained by other enrichment processes such as SNe II.
Therefore, it is highly likely that the $z=3.54$ system is a gas cloud
enriched by SNe Ia. More detailed arguments based on 
abundance estimate using CLOUDY photoionization modeling will be
presented in our separate paper (S. Kondo et al. in preparation).

If the gas cloud was truly enriched by SNIa explosion, the total amount of the iron in the gas should be consistent with the yield of the iron from SNIa explosion.
Assuming the
  \ion{Fe}{2} absorbing gas is in the form of shell with $R=50-100$ pc as
  Equation (\ref{himass}), the \ion{Fe}{2} mass is estimated as 0.07-0.29
  $M_\odot$ from the component 3 of image A. Our
  estimated \ion{Fe}{2} mass is consistent with the observed mass of SNeIa. \citet{sca10} and \citet{sil11} suggest that the estimated mass of radio active $^{56}$Ni (eventually decays to $^{56}$Fe) ejected from an SN Ia ranges from 0.02 $M_\odot$ to 1.7 $M_\odot$. Note that the slightly low estimated energy of the SN (Section 5.2.3) also favors an SNIa interpretation rather than other types of SN with more energetic explosions. Because our estimate does not include the iron in other ionization states, such as \ion{Fe}{3}, within the shell as well as the iron \textit{inside} the shell, the total mass of iron is expected to be more than the estimated value.

\subsubsection{Fe Localization}

\begin{figure}
 \centering
 \includegraphics[width=8cm,clip]{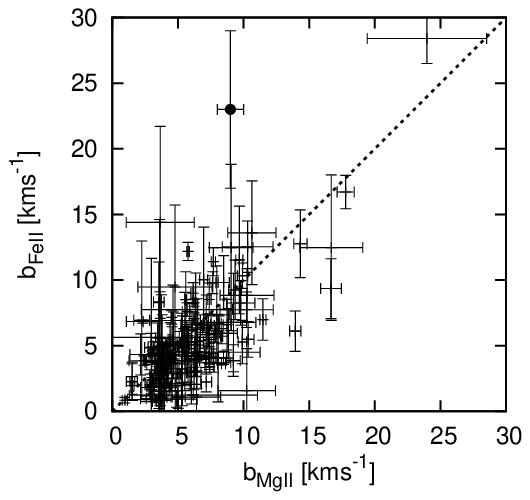}
 \caption{Correlation of Doppler widths of \ion{Mg}{2} and \ion{Fe}{2} absorption lines of weak \ion{Mg}{2} systems. The crosses are from \citet{rig02}, \citet{chu03}, \citet{nar08}, while the filled circle shows the $z=3.54$ system (component 3 of image A). The dotted line shows the location where the widths of \ion{Mg}{2} and \ion{Fe}{2} are identical. While the crosses are distributed along or below the dotted line, the point of the $z=3.54$ system is located above the line with a large offset.}
 \label{dopplar}
\end{figure}

Another characteristic of the \ion{Fe}{2} absorption line is the Doppler width ($23\pm6$ km s$^{-1}$) that appears to be unusually broad compared with that of $\alpha$-element (e.g., $b_{\text{\ion{Mg}{2}}} \sim 9$ km s$^{-1}$).
This is quite strange for
a QSO absorption system, because the mass of an iron atom is much
heavier than that of $\alpha$ elements, thus the width of an \ion{Fe}{2}
absorption line should be smaller than that of $\alpha$ elements. 
We first compared the widths of \ion{Mg}{2} and \ion{Fe}{2} lines of the $z=3.54$ system with the past surveys of weak 
\ion{Mg}{2} systems \citep{rig02,chu03,nar08}.
Figure \ref{dopplar} shows correlation of the Doppler widths of \ion{Mg}{2} and \ion{Fe}{2} lines of weak \ion{Mg}{2} systems. The points should be located below the dotted line ($b_{\text{\ion{Mg}{2}}} = b_{\text{\ion{Fe}{2}}}$) because iron is heavier than magnesium. In fact, most data points from the literatures are distributed along or below the dotted line. However, the $z=3.54$ system is found to be located significantly above the line even considering the uncertainty, suggesting the uniqueness of this system. 

We first checked the possibility of blending of other absorption lines on the \ion{Fe}{2} absorption lines. For example, if
\ion{Mg}{2} absorption systems existed at $z=$3.22 and $z=$2.87, strong metal absorption lines \ion{Mg}{2} $\lambda \lambda$2796, 2803 would blend with the \ion{Fe}{2} $\lambda$2600 ($\lambda =
11804$ \r{A}) and $\lambda$2383 ($\lambda = 10814$\r{A}) lines,
respectively. However, such absorption systems have not been reported in
\citet{son96} and \citet{rau01a}, who detected
many Lyman series lines and \ion{C}{4} absorption lines of B1422+231. Therefore, we conclude that the broad features of \ion{Fe}{2} lines are real.

\begin{figure}
 \centering
 \includegraphics[width=8cm,clip]{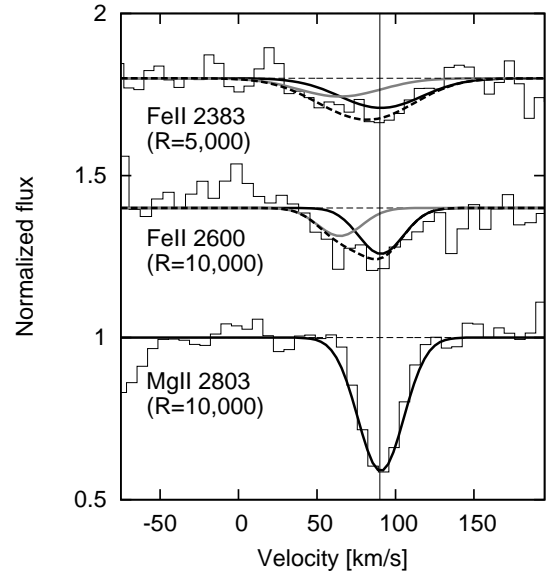}
 \caption{Voigt profile fitting to the \ion{Mg}{2} $\lambda$2803 (bottom), \ion{Fe}{2} $\lambda$2600 (middle) and $\lambda$2383 (top) absorption lines of component 3 of image A. The thin lines show the observed spectra. The solid thick lines show the fitted Voigt profiles with the Doppler width and redshift fixed to the values of \ion{Mg}{2} absorption lines. The gray lines show the fitted Voigt profiles to the residual absorption. The dashed lines show the combined profiles of black and gray components.}
 \label{vpfitcomp}
\end{figure}

Here we try to evaluate the excess \ion{Fe}{2} absorption by decoupling the broad feature into two sub-components. If we compare the \ion{Mg}{2} and \ion{Fe}{2} profiles, the excess exists on the blue side of the \ion{Mg}{2} absorption peak (Figure \ref{Fecomb}). Then, 
we fitted two
velocity components to both \ion{Fe}{2} $\lambda$2600 and $\lambda$2383 lines with
the fixed Doppler widths of 9 km
s$^{-1}$
, which is the upper limit estimated from the Doppler width of \ion{Mg}{2} ($\sim 9$ km s$^{-1}$). The results
are shown in the bottom panel of Figure \ref{vpfitcomp}. The black line
shows the component whose redshift is fixed to the value of the
\ion{Mg}{2} absorption line during the fitting, while the gray line
shows the other excess component, for which the redshift was not fixed. The
 resultant difference of peak velocity between the two velocity components is
$26\pm7$ km s$^{-1}$, and the column densities of blue and red
components are $\log N$ [cm$^{-2}$] = $12.3 \pm 0.2$ and $12.5\pm0.1$,
respectively. This fitting appears to match well with the observed data for
both \ion{Fe}{2} $\lambda$2600 and $\lambda$2383 lines.

\begin{figure}
 \centering
 \includegraphics[width=8cm,clip]{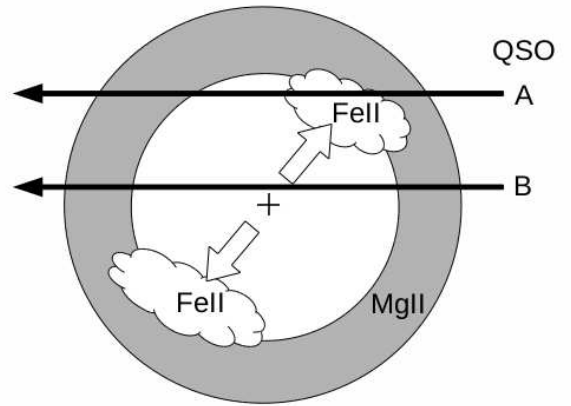}
 \caption{Schematic image of the SNR shell of the $z=3.54$ system. The Fe-rich gas clouds are localized in the expanding shell, while the $\alpha$ element gas clouds, such as \ion{Mg}{2}, are distributed homogenously in the expanding shell.}
 \label{shellfe}
\end{figure}

Now, the question is what is the blue
component ?  Because there is no obviously corresponding
$\alpha$-element absorption lines, this iron gas cloud is inferred to be localized
in the shell.
The localization of iron in the SNR also supports the SNIa origin of
the SNR shell: the \ion{Fe}{2} features imply the ejection of the mass
from the SN Ia, while the other $\alpha$-element absorption
lines are likely to be the interstellar gas scrambled by the shock of the
SNR. In our Galaxy, \citet{ham97} observed SN1006, which is an SN Ia remnant, with absorption lines on a spectrum of Schweizer
and Middleditch star, whose sightline intersects near the center of
SN1006. Despite the age difference ($\sim$1000 yr for SN1006, $\sim 10^5$ yr for the $z=3.54$ system), it would be useful to compare those two objects in detail to
infer the nature of the $z=3.54$ SNR shell because the geometrical configuration is very similar. For SN1006, very broad
($\sim$ 5,000 km s$^{-1}$) \ion{Fe}{2} absorption line is detected
at the systemic velocity while the ejected \ion{Si}{2} shows a
redshift of $\sim$ 5,000 km s$^{-1}$. This suggests that the mixing of
ejected iron with interstellar gas occurs after the mixing of
 ejected $\alpha$-elements. The two velocity sub-components of \ion{Fe}{2}
absorption lines of component 3 of image A can be interpreted as meaning that the red
component with the same velocity as the $\alpha$-element lines
comes from Fe gas that mixed with the scrambled interstellar gas, and
the blue component comes from Fe gas that has not been mixed with it
yet (Figure \ref{shellfe}). Therefore, we might witness the mixing process of ejected
iron with interstellar gas at $z=3.54$.

\subsubsection{SNe Ia at High Redshift}

In this section, we saw the evidence of the richness and localization of
iron in the SNR shell at $z=3.54$. The detected iron richness cannot be
explained by processes other than SN Ia. Moreover, the suggested
localization of iron in the shell supports the SN Ia
interpretation. From these facts, we conclude that the $z=3.54$ system
is an SN Ia remnant.

The extragalactic SN Ia has been studied extensively for the supernova
cosmology \citep{goo11}. However, even the most distant SN Ia event ever
detected is at $z=1.55$ \citep{rod11} and more distant objects that are
important for the study of cosmic chemical enrichment history are hard
to detect and with 8-10 m class telescopes. For studying such high-$z$
SNe Ia, absorption systems toward gravitational-lensed QSO may serve as
good targets as is the case for this $z=3.54$ system toward
B1422+231. Even for the single sightline, the Fe-rich absorption systems
as seen in Figure \ref{narayanan} would become good targets for studying
chemical enrichment history at high-redshift ($z>2$) with more data
available with sensitive high-resolution spectroscopy with adaptive
optics \citep[see][]{kob05}.

\section{Summary}
We obtained near-infrared high-resolution ($R=$10,000) spectra of images
A and B of gravitationally lensed QSO B1422+231 with the Subaru 8.2 m
telescope and the IRCS echelle spectrograph. Although the observed PSFs
of images A and B are partially overlapping in the slit, we managed to
extract spectra for each image A and B to examine the differences of
absorption lines. We detected \ion{Mg}{2} $\lambda\lambda$2796, 2803
absorption lines of the $z=3.54$ system, which had been found in images
A and C in optical spectra by RSB99. Corresponding \ion{Fe}{2}
$\lambda$2383, $\lambda$2600 absorption lines are also detected but only
for the component 3 of image A.  The projected separation between A and
B images at $z=3.54$ is just $8.4h_{70}^{-1}$ pc and we found
considerable differences of column density ($\Delta (\log N) \sim 0.3$
dex) and velocity shear ($\Delta v \sim 10$ kms$^{-1}$) between both
images on such a small scale. These differences suggest the smallest
structure of gas clouds ever detected for QSO absorption systems.

Considering the physical origin of the differences of $\alpha$-elements absorption lines among three images A, B, and C, we conclude that the $z=3.54$ system is an expanding shell as originally suggested by RSB99. The information of \textit{three images} enable us to analyse the 3D structure of the absorbing gas cloud and concluded that this $z=3.54$ system is an SNR shell whose radius and expansion velocity are 50$-$100 pc and $\sim$ 130 km s$^{-1}$, respectively. We estimated several physical parameters of the shell, which are found to be consistent with SNRs: the age is $\sim 10^5$ yr, the mass of the shell is $\sim$ 100 $M_\odot$, and the energy of the SNe is $\sim 10^{50}$ erg. 

We also found that the \ion{Fe}{2} absorption lines of $z=3.54$ system have much larger column density and Doppler width than those of the weak \ion{Mg}{2} systems in the literature. The small column density ratio $\log N$(\ion{Mg}{2})/$N$(\ion{Fe}{2}) (=0.31$\pm$0.07) is indicative of the richness of the iron of the $z=3.54$ system. We also estimated the mass of \ion{Fe}{2} as 0.07$-$0.29 $M_\odot$ which is roughly consistent with the yield of observed SNe Ia. Moreover, the large Doppler width of \ion{Fe}{2}, which is interpreted as the existence of one more velocity component that is blueshifted and extremely Fe-rich, suggests that the Fe-rich gas cloud is localized in the expanding shell of SNR. From these results, we conclude that the SNR shell at $z=3.54$ is produced by an SN Ia explosion.

Supernova explosions are thought to be one of the most important processes that drive the formation of galaxy through the dynamical energy input and the chemical enrichment. If this $z=3.54$ system is truly an SNR, it is the farthest sample of a supernova ever observed. 
If higher S/N and/or higher spectral resolution data of B1422+231 were obtained in the future, the difference between iron and $\alpha$-elements would be more clearly confirmed. Then, physics of the mixing of interstellar gas with ejected gas from supernovae could be discussed in detail. If the spectrum of image D is also obtained, the radius and expansion velocity can be strictly determined with the expanding shell model.

This $z=3.54$ system illustrates the power of the gravitational lensing effect for the study of QSO absorption systems with the extremely high spatial resolution and multiple sightlines. In the case of present study, the separation between images A and B is only $<$ 10 pc at z=3.5 that corresponds to about 1 mas angular resolution. More observations of gravitationally lensed QSOs will reveal the kinematics (e.g., expansion) of gas clouds on interstellar scales at high redshift during crucial phases of galaxy formation.

\ 

We are grateful to all of the IRCS and AO team members and the Subaru
Telescope observing staffs for their efforts, which made it possible to
obtain these data. We are grateful to Dr. Rauch for kindly providing us
their B1422+231 optical spectra obtained with Keck/HIRES. This work was
supported by KAKENHI Grant-in-Aid for Scientific Research(B)
(No. 20340042; N. Kobayashi) from JSPS, and in part by the Graduate
University for Advanced Studies (Sokendai). This research has been
partially supported by the Private University Strategic Research
Foundation Support Program of the Ministry of Education, Science, Sports
and Culture of Japan, S0801061.

%We are grateful to all of the IRCS and AO team members and the Subaru Telescope observing staffs for their efforts, which made it possible to obtain these data. We are grateful to Dr. Rauch for kindly providing us their B1422+231 optical spectra obtained with Keck/HIRES. This work was supported in part by The Graduate University for Advanced Studies (Sokendai).

%[7]

\end{document}